\documentclass[12pt,preprint]{aastex}

\newcommand{\etal}{et al.~}

\shorttitle{The Globular Cluster System of M60. I.}
\shortauthors{Lee et al.}

\begin{document}


\title{The Globular Cluster System of M60 (NGC~4649).
I. CFHT MOS Spectroscopy and Database\altaffilmark{*}}

\author{Myung Gyoon Lee\altaffilmark{1,**}, Ho Seong Hwang\altaffilmark{1},
Hong Soo Park\altaffilmark{1}, Jang-Hyun Park\altaffilmark{2}, \\
Sang Chul Kim\altaffilmark{2}, Young-Jong Sohn\altaffilmark{3}, 
Sang-Gak Lee\altaffilmark{1}, Soo-Chang Rey\altaffilmark{4}, \\
Young-Wook Lee\altaffilmark{3},
Ho-Il Kim\altaffilmark{2}} 

\altaffiltext{1}{Astronomy Program, Department of Physics and Astronomy,
Seoul National University, 56-1 Sillim 9-dong, Gwanak-gu, Seoul 151-742, Korea}

\altaffiltext{2}{Korea Astronomy and Space Science Institute,
Daejeon 305-348, Korea}

\altaffiltext{3}{Center for Space Astrophysics, Yonsei University,
Seoul 120-749, Korea}

\altaffiltext{4}{Department of Astronomy and Space Sciences,
Chungnam National University, Daejeon 305-764, Korea}

\altaffiltext{5}{*Based on observations obtained at the Canada-France-Hawaii
Telescope (CFHT) which is operated by the National Research
Council of Canada, the Institut National des Science de l'Univers
of the Centre National de la Recherche Scientifique of France,
and the University of Hawaii.
This work is organized by the Korea Astronomy and Space Science Institute.}

\altaffiltext{6}{**mglee@astro.snu.ac.kr}


\begin{abstract}
We present the measurement of radial velocities for globular clusters in M60,
giant elliptical galaxy in the Virgo cluster.
Target globular cluster candidates were selected using the Washington photometry
based on the deep $16\arcmin \times 16\arcmin$ images taken at the KPNO 4m
and using the $VI$ photometry derived from the HST/WFPC2 archive images.
The spectra of the target objects were obtained using
the Multi-Object Spectrograph (MOS) at the Canada-France-Hawaii Telescope (CFHT).
We have measured the radial velocity for 111 objects in the field of M60:
93 globular clusters
(72 blue globular clusters with $1.0\le(C-T_1)<1.7$ and 21 red globular clusters with  $1.7\le(C-T_1)<2.4$ ),
11 foreground stars, 6 small galaxies,
and the nucleus of M60.
The measured velocities of the 93 globular clusters range from $\sim 500$ km~s$^{-1}$ to $\sim 1600$ km~s$^{-1}$,
with a mean value of $1070_{-25}^{+27}$ km~s$^{-1}$,
which is in good agreement with the velocity of the nucleus of M60 ($v_{\rm gal}=1056$ km~s$^{-1}$).
Combining our results with data in the literature, 
we present a master catalog of radial velocities for 121 globular clusters in M60.
The velocity dispersion of the globular clusters in the master catalog
is found to be
$234_{-14}^{+13}$ km~s$^{-1}$ for the entire sample,
$223_{-16}^{+13}$ km~s$^{-1}$ for 83 blue globular clusters, and
$258_{-31}^{+21}$ km~s$^{-1}$ for 38 red globular clusters.

\end{abstract}


\keywords{galaxies: clusters: general --- galaxies: individual
(M60, NGC 4647) --- galaxies: kinematics and dynamics --- galaxies: star
clusters --- galaxies:spectroscopy}

\section{INTRODUCTION}

Globular clusters  are among the oldest stellar systems in galaxies so
they provide important clues for understanding the formation and
early evolution of their host galaxies as well as themselves.
Globular clusters are distributed in a much wider region than the halo stars in
their host galaxy,
and thousands of them are found in giant elliptical galaxies (gEs).
Thus globular clusters are a very efficient tool to study the structure and kinematics
of both the outer halo  and the inner region of nearby gEs.

To date numerous studies were devoted to investigating the photometric properties of the
globular clusters in gEs (e.g., see the reviews by \citealt{lee03,bro06}).
However, much less is known about the kinematics of the globular clusters in gEs,
because it was difficult to obtain spectra of a larger number of globular clusters in gEs
until the multi-object spectrograph became available at the large telescopes.

Recently several studies of the kinematics of the globular clusters in gEs came out,
but the number of gEs covered in these studies is still small:
M49 (NGC 4472) \citep{zep00, cot03}, M87 (NGC 4486) \citep{coh97, kis98b, cot01},
NGC 1399 \citep{kis98, min98, kis99, ric04}, NGC 4636 \citep{sch06},
M60 (NGC 4649) \citep{pie06, bri06}, and NGC 5128 \citep{pen04a, pen04b, woo07}.
One notable feature revealed by these studies is that the globular cluster systems
in these gEs show diverse kinematics in velocity dispersion, rotation, and
their radial variation, 
making it difficult to draw any strong conclusion on the uniform formation
history of the globular cluster system in gEs. Therefore, it is needed to
investigate more gEs to understand the diversity of kinematics of these globular
cluster systems.

We have been carrying out a study of kinematics of the globular clusters in M60.
In this paper we present the measurement of radial velocities for the globular clusters in M60,
and will present the dynamical analysis of the data in the following paper \citep{hwa07}.
M60 is a giant elliptical galaxy in the Virgo cluster,
being slightly less luminous ($M_V=-22.44$) than the brightest Virgo galaxies
M87 ($M_V=-22.62$ mag) and M49 ($M_V=-22.83$ mag).
We adopted a distance to M60, 
17.3 Mpc ($(m-M)_0=31.19\pm 0.07$) 
based on the surface brightness fluctuation method in \citet{mei07}
for which one arcsec corresponds to 84 pc. 
Foreground reddening toward M60 is very small, $E(B-V)=0.026$ \citep{sch98},
corresponding to $E(C-T_1)=1.966 E(B-V)=0.051$ and $A(T_1)=0.071$ and $A(V)=0.088$.
Effective radius, ellipticity, and position angle of M60 are 
$R_{eff}=9.23$ kpc, $\epsilon = 0.21$, and PA$=105$ deg, 
respectively \citep{lee07}.

M60 has a companion spiral galaxy, NGC 4647 (SAB(rs)c),
located  $2.\arcmin5$ from the center of M60 in the north-west
direction ($\sim$12.6 kpc for a distance to M60).
While \citet{san94} described that NGC 4647 is not interacting with M60,
recent observational results show some evidence of interaction between the two:
(1) the morphology of NGC 4647 is clearly asymmetric \citep{koo01};
(2) the inner region of M60 has a strong rotational support in contrast to other
giant elliptical galaxies, and has an asymmetric rotation curve \citep{pin03,deb01};
(3) there is seen a filament extending to the
northeastern edge of M60 in the X-ray image \citep{ran06}; and
(4) there are a large number of blue luminous star clusters in NGC 4647 that were probably formed during the interaction between the two galaxies \citep{lee07}.

Thera are several photometric studies of the globular clusters in M60.
\citet{cou91} presented the first
photometric study based on the CCD imaging
of the globular cluster system in a small
$2\arcmin.1 \times 3\arcmin.4$ field of M60,
finding a large dispersion in color distribution, and a radial gradient in
the mean cluster colors.
\citet{har91} derived from the same data the $B$ band
luminosity function up to $B\sim$26 mag of the globular clusters, 
and found that the globular clusters follow more extended distribution than the stellar light.
Later, photometric studies based on the HST images and ground-based images
revealed that the M60 globular cluster system has a distinct bimodality in
the color distribution \citep{nei99,kun01,lar01,for04,pen06, lee07}.
\citet{for04} found from the analysis of wide field images
obtained using Gemini/GMOS
that the red globular clusters have a similar surface density
distribution to that of stellar light, which is steeper than that
of the blue globular clusters. 
Recently, it is found that M60 is one of the galaxies that show ``blue-tilt''
in the color-magnitude diagram of globular clusters 
(\citealt{str06,mie06,lee07}). 
Recent studies based on 
the {\it Chandra} data found that roughly 47\% of the X-ray discrete
sources are identified with globular clusters \citep{sar03,ran04}.
By cross-correlating {\it Chandra} point sources and optical globular cluster candidates,
\citet{kim06} found that the mean probability for a  globular cluster to harbor a
low mass X-ray binary in M60 is about 6.1$\pm$0.9\%, and that
probability for the red globular clusters is larger than that for the blue globular clusters.

There are only two previous studies on the spectroscopy of the globular clusters in M60. 
\citet{pie06} published a spectroscopic study of 38 globular clusters in M60 based on the data obtained using the Gemini/GMOS.
They derived ages and metallicities of the globular clusters from the spectral line indices.  They found no obvious signs of a recent starburst, interaction or merger in their estimates of the ages and metallicities. 
\citet{bri06} presented a kinematic study of the M60 globular cluster system using the same spectral data as used by \citet{pie06}. They found that the velocity dispersion of the blue globular clusters
is smaller than that of the red globular clusters, and that there is no, if any,
rotation of the globular cluster system in M60.

In this paper
we present the measurement of radial velocities for the globular clusters in M60
using the spectra obtained at the 3.6m Canada-France-Hawaii Telescope (CFHT), 
and the kinematic analysis of the data
will be given in the companion paper \citep{hwa07}.
This paper is composed as follows.
Section 2 describes the selection of the targets for spectroscopy,
and Section 3 describes spectroscopic observation.
Section 4 explains the reduction of the data, and Section 5 presents
the final data of radial velocities for globular clusters and other objects
in the field of M60.
Primary results are summarized in the final section.

\section{SPECTROSCOPIC TARGET SELECTION} 

We selected globular cluster candidates using two sets of optical images of M60:
(a) F555W ($V$) and F814W ($I$) images of the inner region of M60 in the HST/WFPC2 archive
(P.I.: J. R. Westphal, Program ID: 6286) that were also used in the photometric analysis by \citet{nei99, kun01, lar01},
and
(b) Washington $CT_1$ images of a $16\arcmin\times16\arcmin$ field including the center of M60
obtained at the KPNO 4m telescope that were used in the photometric study of globular clusters in M60 by \citet{lee07}. 
\citet{lee07} derived, 
from the comparison of ($C-T_1$) colors and ($V-I$) colors for
the objects with $T_1<22$ mag common between the KPNO images
and HST WFPC2 images of M60,
a linear relation for the color range of $0.9<(C-T_1)<2.4$:
 $(V-I)=0.407 (C-T_1) +0.459$ with rms=0.047  for 72 objects.
We converted $(V-I)$ colors of the objects detected in the HST/WFPC2 field
into  $(C-T_1)$ colors using this transformation equation. %

It is known that most bright globular clusters in gEs have colors $1.0\le(C-T_1)<2.4$
(or $0.87< (V-I)<1.44$, \citealt{lee07}).
We selected as the spectroscopic targets
the bright globular cluster candidates with $1.0\le(C-T_1)<2.4$ and $19<T_1<22$ mag.
We included also several small galaxies, a few red objects ($(C-T_1)>2.4$), 
and the nuclei of M60 and its companion spiral galaxy NGC 4647.
The number of the targets in the final sample 
is 165.

\section{OBSERVATION} 

We used 
the Multi Object Spectrograph (MOS) \citep{lef94} 
at the CFHT during two observing runs
in February 2002 and May 2003 to obtain the spectra of the targets.
Table 1 summarizes the information on our spectroscopic setup and
Table 2 lists the log of our observations.

One mask for the MOS covers a field of $9\arcmin.3 \times 8\arcmin.3$.
We observed six mask fields inside the $16\arcmin \times 16 \arcmin$ field,  as marked in Figure 1.
Figure 1 displays the grayscale map of the $T_1$ image of M60 taken at the KPNO 4m \citep{lee07}. 
We created first the model image of the galaxy halo light using the ellipse fitting task
IRAF/ELLIPSE\footnotemark\footnotetext
{IRAF is distributed by the
National Optical Astronomy Observatories, which are operated by the
Association of Universities for Research in Astronomy, Inc., under
contract to the National Science Foundation.}.
Then the model image of the halo light of M60 was subtracted
from the original image to show better the globular
cluster candidates in M60, as seen in Figure 1.

We used a slit length of 10$\arcsec$ 
 to cover both the target and its sky background, 
 and a slit width of 1.5$\arcsec$ 
 which is appropriate for the seeing condition.
The seeing during the observation was about one arcsec.
Slits in the mask were cut using the LAMA machine in the mountain,
and the number of slits in one mask ranges from 21  to 34.
We used the grism, B600, giving a dispersion of 1.44 \AA~pixel$^{-1}$
and a spectral resolution of about 3.5 \AA.
The resulting spectral coverage is $3600-7000$ \AA.

We obtained three 2400 s exposures for targets in all but one mask field
(two exposures for M60$\_$HW1 mask), with comparison spectra for
Ne-Ar-Hg lamps taken before and/or after each exposure.
For the calibration of the radial velocity, we obtained long slit spectra of three
Galactic globular clusters (NGC 2419, NGC 6341, and NGC 6356).
Table~3 lists the metallicity and radial velocity for these clusters
(\citet{har96} and references therein).
We selected two metal poor globular clusters
(NGC 2419 with [Fe/H]=$-2.12$ dex
and NGC 6341 with [Fe/H]=$-2.28$ dex)
and one metal-rich globular cluster (NGC 6356 with [Fe/H]=$-0.50$ dex)
to cover a wide range of metallicity.

\section{DATA REDUCTION} 

\subsection{Spectroscopic Data Reduction}

We have reduced the spectroscopic data using the IRAF software.
After basic preprocessing of the raw spectral images,
we removed the cosmic rays in each image using
IRAF/COSMICRAYS task, and combined three exposure images
of each mask.
The spectra of the targets from the combined images
were traced, extracted, and sky-subtracted using the IRAF/APALL task.
We could not extract spectra of some faint targets
because of low signal-to-noise ratio.

We identified about 30 emission lines
in the comparison lamp spectra from 4000 \AA~to 7000 \AA,
and used them for the wavelength calibration.
The typical rms error of the wavelength calibration is 0.1$-$0.3 \AA.

Figure 2 displays the example spectra of five objects:
a blue globular cluster in M60 (ID 151 with $T_1=20.22$ and $C-T_1=1.32$),
a red globular cluster in M60 (ID 146 with $T_1=20.23$ and $C-T_1=1.77$),
two Galactic globular clusters (NGC 2419 and NGC 6356)
and the nucleus of M60.
Several typical absorption features for Galactic globular clusters are seen
in the spectra of globular clusters in M60. The blue continuum of the red globular
cluster in M60 is much steeper than that of the blue globular cluster in M60,
as seen also in those of the Galactic globular clusters (NGC 6356 and NGC 2419).
Absorption features in the spectrum of the M60 nucleus are seen to be
much broader than those in the globular clusters.

\subsection{Radial Velocity Determination}

We determined the radial velocity for the targets
using the Fourier cross-correlation task, IRAF/FXCOR \citep{ton79}.
We fitted the continuum of the spectra using the spline-fit
with $2\sigma$ clipping for low level and $4\sigma$ clipping for
high level, and subtracted the resulting fit from the original spectra.
We used the resulting spectra to measure the radial velocity with FXCOR.

After several tests, we decided to use the wavelength
range $3900-5500$ \AA~  for cross-correlation.
We measured the radial velocity for the targets,
using three templates of Galactic globular clusters
(NGC 2419, NGC 6341 and NGC 6356), and took the error-weighted average of the three
measurements to get the final value for each target.
The error of the measured radial velocity is then 
  $<\epsilon_v>=(\Sigma\epsilon_i^{-2})^{-1/2}$.

Among the spectra of 165 targets 
we could not determine the radial velocities for 54 objects because
of poor quality of the spectra. 
The final number of targets 
for which we determined the radial velocity is 111:
93 globular clusters, 11 foreground stars, 6 small galaxies, and the nucleus of
M60.

Figure 3 displays the errors of the measured radial velocities
  versus $T_1$ magnitudes and galactocentric distances of the 110 targets,
  excluding the nucleus of M60.
The errors for our measurements of radial velocities
of the targets range mostly from 20 km s$^{-1}$ to 110 km s$^{-1}$
with a mean of $54\pm22$ km s$^{-1}$.
The mean errors increase with increasing $T_1$ magnitude
($44\pm 18$ km s$^{-1}$ for the bright objects with $T_1<20.7$ mag
and $61\pm 29$ km s$^{-1}$ for the faint objects with $T_1>20.7$ mag in the inner region),
while they change little depending on the galactocentric distance
($53\pm 26$ km s$^{-1}$ for 38 objects in the inner region ($R<200\arcsec$),
and $57\pm 24$ km s$^{-1}$ for 55 objects in the outer region ($R>200\arcsec$)).
The mean error for the red objects, $46\pm 11$ km s$^{-1}$,
is slightly smaller than that for the blue objects, $58\pm 25$ km s$^{-1}$.

\subsection{Comparison with Previous Studies}

There is only one previous spectroscopic study of the globular clusters in M60,  given by \citet{pie06}.
\citet{pie06} published a catalog of spectroscopic data including radial velocities for 38 globular clusters in M60
based on the spectra obtained using the Gemini/GMOS.
There are nine objects in common between this study and
\citet{pie06}.
We have compared our velocity measurements with their measurements
for these common objects, as displayed in Figure 4.
It shows a good agreement between the two studies
except one object (ID 230 in this study and 1574 in \citealt{pie06}).
Our value for ID 230 object is $v_p = 1480\pm 22$ km~s$^{-1}$, which is about twice larger than
the value given by \citet{pie06}, $v_p = 703\pm 53$ km~s$^{-1}$.
We checked the image of ID 230, finding no nearby neighbor within the radius of 5.8 arcsec from
this object. Therefore there is no possibility of misidentification of this object. The cause for this large difference for ID 230 is not known.
From the weighted linear fit discarding ID 230,
  we derive a transformation relation between the two measurements,

\begin{equation}
v_{\rm This~ study} = 1.04 (\pm0.05)~v_{\rm Pierce ~et ~al.} -67.9 (\pm52.5) {\rm~km~s}^{-1}. \label{eq-trans}
\end{equation}

We have derived the radial velocity for the nucleus of M60 as $v_p = 1056\pm64 $ km~s$^{-1}$,
  which is consistent with that of $v_p = 1117\pm6$ km~s$^{-1}$ for M60 \citep{gon93}. 
The larger error in our estimate of the M60 systemic velocity
is primarily due to the fact that we used, as templates, the spectra
of GCs with small velocity dispersion to estimate the velocity of
the nucleus of M60 that has much larger velocity dispersion.

\section{RESULTS} 

We have produced a master catalog of radial velocities for the objects in M60,
combining the catalog in this study with the catalog of 38 globular clusters in M60 given by \citet{pie06}.
The total number of the objects in the master catalog is 139,
excluding the nucleus of M60.
We transformed the radial velocities given by \citet{pie06} to our system,
using the transformation equation (1) 
for the following analysis.

\subsection{Velocity Distribution and Membership}

We have determined the membership of the objects in the master catalog
using the distribution of radial velocity.
In Figure \ref{fig05}, we plotted ($C-T_1$) colors versus the radial velocities (the lower left panel),
  the radial velocity distribution (the upper panels), and the $(C-T_1)$ color distribution
  (the lower right panel) for the 139 objects. We also plotted in the lower right panel,
  the color distribution of the photometric globular cluster candidates given in \citet{lee07}. 
\citet{lee07} found that the color distribution of the globular clusters is clearly bimodal
  through the KMM mixture modeling routine \citep{ash94}, 
  and is well fit by two Gaussians with peaks at $(C-T_1)=1.37$ ($\sigma=0.16$) and 1.87 ($\sigma=0.23$).
The minimum between the two components is at $(C-T_1)=1.7$, which was used
for diving the entire sample into blue globular clusters and red globular clusters.

The radial velocity distribution of the 139 objects is approximately Gaussian
centered around the radial velocity for the M60 nucleus, $v_p\approx 1056$ km~s$^{-1}$, 
with a weak excess around the zero velocity.
We selected, as genuine M60 globular clusters,
those of which radial velocities are in the range of $500 \le v_p \le 1600~{\rm
km~s}^{-1} $ and ($C-T_1$) colors are in the range of $1.0 \le (C-T_1) < 2.4$.
The dashed-line box in Figure 5 represents the adopted selection criteria
for ($C-T_1$) colors and radial velocities.
The selection color range is equivalent to a metallicity range of
$-$2.01$\lesssim$~[Fe/H]~$\lesssim$0.91 dex,
$-$2.08$\lesssim$~[Fe/H]~$\lesssim$1.48 dex,
$-$2.06$\lesssim$~[Fe/H]~$\lesssim$0.21 dex, and
$-$1.95$\lesssim$~[Fe/H]~$\lesssim$0.69 dex
using the color-metallicity relations of \citet{lee07} (double linear and 3rd order polynomial equations),
\citet{har02}, and \citet{coh03}, respectively.
The total number of the genuine globular clusters of M60  in this catalog is 121.
There are 83 blue globular clusters with $1.0 \le (C-T_1) < 1.7$
and 38 red globular clusters with $1.7\le(C-T_1) < 2.4$.

There are six small galaxies in the list of our measurement
of the radial velocity:
IDs 176, 444, 9930, 9988, 99213 and 99214.
They appear as small extended objects in the KPNO images.
Two of these are previously known galaxies:ID 444 = VCC1963 and ID 99214 = VCC1982
\citep{bin85}. 
ID 444 (VCC1963) is a low surface brightness galaxy with a bright nuclues.
Only one (ID 99213) among these has a radial velocity larger than the upper boundary for the globular clusters in M60.

There are 12 objects with radial velocities smaller than the lower boundary for the globular clusters.
We consider these objects as foreground stars.
One object (ID 661 in this study, and ID 412 in \citealt{pie06}) among these has a radial
velocity, $v_p = 441 \pm 50$ km~s$^{-1}$ (originally $v_p = 483 \pm 47$ km~s$^{-1}$ in \citealt{pie06}),
slightly smaller than the lower boundary for the globular clusters.
This object may be a globular cluster of M60, but we put this into the class of foreground
stars according to the boundaries for the globular clusters we adopted.

Among the 110 objects with radial velocities derived in this study,
93 targets are found to be genuine globular clusters in M60 and
10 targets turned out to be foreground stars.
The remaining 7 targets that were included as non-globular clusters turned out to be one foreground star and
six galaxies. 
Therefore the success rate of photometric searching for globular clusters,
becomes about 90\%,
showing that the method to select the globular cluster candidates using the $(C-T_1)$
color and the morphological classifier is very efficient (see \citet{lee07} for more details).

In Table \ref{tab-m60gc}, we list the photometric and kinematic
data set for all 139 spectroscopic targets with measured radial velocity.
The first column represents
identification numbers. The second and third columns give, respectively,
the right ascension and the declination (J2000). The magnitude and
color information in columns 4 and 5 are taken from \citet{lee07}. 
The sixth column gives the radial velocity and its error measured in this study.
The seventh and eighth columns give the radial velocity and its error measured 
( transformed into our velocity system) in \citet{pie06}, and their IDs, respectively.
The final column gives a merged, weighted mean velocity for all GCs.
The genuine M60 globular clusters
are listed first
in the table, followed by foreground stars 
and small galaxies.

\subsection{The Final Sample of Globular Clusters in M60}

The positions of the genuine globular clusters are marked in Figure 1.
These clusters are located in the range of projected galactocentric distance 
30 to 540 arcsec 
(corresponding to 2.52 to 45.36 kpc and 0.27 to 4.91 $R/R_{eff}$).
The velocity dispersion 
(the biweight scale of \citealt{beers90a}) is estimated
to be $\sigma_p=239_{-16}^{+12}$ km~s$^{-1}$ for the 93 clusters measured in this study,
and $\sigma_p=234_{-14}^{+13}$ km~s$^{-1}$ for all 121 clusters in the master catalog.
The velocity dispersion of all 83 blue globular clusters is derived to be
$\sigma_p=223_{-16}^{+13}$ km~s$^{-1}$, which is marginally smaller than that of
the 38 red globular clusters ($\sigma_p=258_{-31}^{+21}$ km~s$^{-1}$).
The mean value of the radial velocity  
(the biweight location of \citealt{beers90a})
is $\overline{v_p}=1070_{-25}^{+27}$ km~s$^{-1}$ for the 93 clusters measured in this study,
and $\overline{v_p}=1073_{-22}^{+22}$ km~s$^{-1}$ for all 121 clusters in the master catalog.
This value is very similar to the systemic velocity of M60, $v_{\rm gal}=1056\pm64$ km s$^{-1}$.

Figure~\ref{fig06} displays the $T_1 - (C-T_1)$ color magnitude diagram and color
distribution of 121 genuine globular clusters in M60 globular clusters in comparison
with other photometric candidates given in \citet{lee07}. 
The genuine globular clusters in the master catalog have magnitudes $19.7 < T_1 < 22.0$,
belonging to the brightest population in M60.
The genuine globular clusters show a broad peak covering
a range of $1.1<(C-T_1) < 2.2$,
while the photometric globular cluster candidates
show a clear bimodality with two peaks of blue and red globular clusters.

In Figure 7 we plot radial velocities versus $T_1$ magnitudes and
$(C-T_1)$ colors of the genuine globular clusters of M60.
There is seen little systematic dependence of the radial velocity of the globular clusters
either on magnitude or color.
In Figure 8 we display radial velocities versus projected galactocentric
radius $R$ [arcsec] and position angle $\Theta$ [deg]
(measured to the east from the north) of these clusters.
It is seen that the velocity dispersion of the clusters changes little
with the galactocentric radius.
However, it appears that the velocity for the globular clusters may change sinusoidally
depending on the position angle, indicating a hint for rotation of the
globular cluster system of M60.
A full dynamical analysis of the kinematics of the globular cluster system of M60
using these data will be given in the following paper \citep{hwa07}.

\section{SUMMARY}

We present the measurement of radial velocities for 111 objects in the field of M60,
giant elliptical galaxy in the Virgo cluster, using the spectra obtained with
the MOS at the CFHT. Primary results of this study are summarized below.

\begin{enumerate}

 \item We have measured the radial velocities of 111 objects in the field of M60:
93 globular clusters (72 blue globular clusters with $1.0\le(C-T_1)<1.7$ and 21 red globular clusters with  $1.7\le(C-T_1)<2.4$),
11 foreground stars, 6 small galaxies, and the nucleus of M60.

\item The mean value and dispersion of radial velocities for the 93 genuine globular clusters
are derived to be $\overline{v_p}=1070_{-25}^{+27}$ km~s$^{-1}$ and $\sigma_p=239_{-16}^{+12}$ km~s$^{-1}$, respectively.

\item We combined our result with the data for 38 globular clusters of M60 given by \citet{pie06},
and created a master catalog of radial velocity for 139 objects in the field of M60.
The number of the genuine globular clusters in M60 in the master catalog is 121, including 83 blue globular clusters and 38 red globular clusters.

\item The mean value and dispersion of radial velocities for the 121 genuine globular clusters
are derived to be $\overline{v_p}=1073_{-22}^{+22}$ km~s$^{-1}$ and $\sigma_p=234_{-14}^{+13}$ km~s$^{-1}$, respectively.
The velocity dispersions of the 83 blue globular clusters and 38 red globular clusters
are estimated to be $\sigma_p=223_{-16}^{+13}$ km~s$^{-1}$ and $\sigma_p=258_{-31}^{+21}$ km~s$^{-1}$, respectively.

\item The velocity dispersion of the clusters changes little with the galactocentric radius.
However, it appears that the velocity for the globular clusters may change sinusoidally
depending on the position angle, indicating a hint for rotation of the
globular cluster system of M60.

\end{enumerate}


\acknowledgments
The authors are grateful to the anonymous referee for constructive comments that helped to improve the manuscript, 
and to the staff members of the CFHT for their warm support during our observations.
M.G.L. is in part supported by the ABRL (R14-2002-058-01000-0).

{\it Facilities:}\facility{CFHT} 




\clearpage

\begin{deluxetable}{lc}
\tablewidth{0pc}
\tablecaption{Information of the Spectroscopic Setup\label{tab-run}}
\tablehead{
\colhead{Item} &
\colhead{Attribute}
}
\startdata
Telescope/Instrument & 3.58m CFHT/MOS \nl
Detector & 2K$\times$4K EEV 1 \nl
grism & B600 \nl
Dispersion & 1.44 \AA \  $\mathrm{pixel^{-1}}$ \nl
Wavelength Coverage & 3600$-$7000 \AA \nl
\enddata
\end{deluxetable}


\begin{deluxetable}{rlccccc}
\tabletypesize{\scriptsize}
\tablewidth{0pc}
\tablecaption{Observing Log for the CFHT MOS Run\label{tab-mask}}
\tablehead{
\colhead{Obs. Date (UT)} &
\colhead{Mask Name} &
\colhead{R.A.(J2000)} &
\colhead{Decl.(J2000)} &
\colhead{Objects/Mask} &
\colhead{T(exp)} &
\colhead{seeing (\arcsec)}
}
\startdata
2002 Feb. ~16 & M60\_SE & 12 43 51.2 & 11 28 18.1 & 21 & 3$\times$2400 s & 1.1 \nl
     Feb. ~17 & M60\_C1 & 12 43 40.0 & 11 32 23.3 & 34 & 3$\times$2400 s & 1.1\nl
     Feb. ~17 & M60\_W1 & 12 43 25.9 & 11 31 14.9 & 30 & 3$\times$2400 s & 0.9 \nl
2003 May 1$-$2 & M60\_HNW & 12 43 28.8 & 11 38 54.4 & 31 & 3$\times$2400 s & 0.7 \nl
     May ~2~ & M60\_HC1 & 12 43 40.0 & 11 33 48.0 & 32 & 3$\times$2400 s & 0.8 \nl
     May ~2~ & M60\_HW1 & 12 43 28.1 & 11 33 01.3 & 32 & 2$\times$2400 s & 0.9 \nl
\enddata
\end{deluxetable}


\begin{deluxetable}{crr}
\tablewidth{0pc}
\tablecaption{Galactic Globular Clusters Used as Templates \label{tab-galgc}}
\tablehead{
\colhead{Cluster} &
\colhead{[Fe/H]\tablenotemark{a}} &
\colhead{$v_{hel}$\tablenotemark{b}} \nl
\colhead{} &
\colhead{(dex)} &
\colhead{(km s$^{-1}$)}
}
\startdata
NGC 6341 (M92) & $-$2.28 & $-$120.3$\pm$0.1 \nl
NGC 2419 & $-$2.12 &  $-$20.0$\pm$0.8 \nl
NGC 6356 & $-$0.50 &   27.0$\pm$4.3 \nl
\enddata
\tablenotetext{a~}{NGC 6341 \citep{beers90b,pet90,sne91,arm94,she96},
NGC 2419 \citep{zinn85,sun88}, and NGC 6356 \citep{zinn85,az88,min95b}}
\tablenotetext{b~}{NGC 6341 \citep{web81,beers90b,cm97,sod99},
NGC 2419 \citep{web81,pet85,poa86,ops93}, and NGC 6356 \citep{web81,zw84,hsm86,az88,min95a}}
\end{deluxetable}
\clearpage


\begin{deluxetable}{rrrrrrrrrrr}
\rotate
\tabletypesize{\scriptsize}
\tablewidth{0pc}
\tablecaption{Radial Velocities of the Objects in the Field of M60 \label{tab-m60gc}}
\tablehead{
\colhead{ID} &
\colhead{R.A.} &
\colhead{Decl.} &
\colhead{$R$} &
\colhead{$\Theta$} &
\colhead{$T_1$} &
\colhead{$(C-T_1)$} &
\colhead{$v_p$} &
\colhead{$v_p$(P06)} &
\colhead{ID(P06)} &
\colhead{$<v_p>$} \nl
\colhead{} &
\colhead{(J2000)} &
\colhead{(J2000)} &
\colhead{(arcsec)} &
\colhead{(deg)} &
\colhead{(mag)} &
\colhead{(mag)} &
\colhead{(km s$^{-1}$)} &
\colhead{(km s$^{-1}$)} & &
\colhead{(km s$^{-1}$)}
}
\startdata

\multicolumn{11}{c}{}\nl \multicolumn{11}{c}{\underbar{Globular
Clusters}}\nl \multicolumn{11}{c}{}\nl
  301 & 12:43:09.37 & 11:28:23.9 & 532.5 & 237.4 & 20.79$\pm$0.04 & 1.18$\pm$0.05 &   941$\pm$ 81 &  ... &   ... &   941$\pm$ 81 \nl
  447 & 12:43:10.99 & 11:30:19.0 & 458.3 & 248.1 & 21.11$\pm$0.01 & 1.29$\pm$0.02 &  1043$\pm$ 43 &  ... &   ... &  1043$\pm$ 43 \nl
  545 & 12:43:11.00 & 11:30:12.1 & 460.8 & 247.3 & 21.29$\pm$0.01 & 1.16$\pm$0.02 &  1261$\pm$ 55 &  ... &   ... &  1261$\pm$ 55 \nl
  157 & 12:43:13.30 & 11:30:04.1 & 433.1 & 244.6 & 20.27$\pm$0.01 & 1.51$\pm$0.01 &   954$\pm$ 57 &  ... &   ... &   954$\pm$ 57 \nl
  151 & 12:43:14.27 & 11:38:06.7 & 480.2 & 308.1 & 20.22$\pm$0.01 & 1.32$\pm$0.01 &  1362$\pm$ 35 &  ... &   ... &  1362$\pm$ 35 \nl
  226 & 12:43:14.31 & 11:32:09.5 & 381.3 & 260.8 & 20.58$\pm$0.02 & 1.64$\pm$0.02 &   601$\pm$ 71 &  ... &   ... &   601$\pm$ 71 \nl
  139 & 12:43:16.04 & 11:30:42.1 & 380.9 & 247.2 & 20.17$\pm$0.02 & 1.72$\pm$0.02 &  1256$\pm$ 31 &  ... &   ... &  1256$\pm$ 31 \nl
  402 & 12:43:17.66 & 11:39:44.9 & 513.4 & 320.2 & 21.01$\pm$0.01 & 1.47$\pm$0.02 &   781$\pm$ 59 &  ... &   ... &   781$\pm$ 59 \nl
  204 & 12:43:18.55 & 11:37:45.8 & 418.3 & 311.2 & 20.50$\pm$0.02 & 1.57$\pm$0.02 &  1337$\pm$ 26 &  ... &   ... &  1337$\pm$ 26 \nl
  246 & 12:43:19.91 & 11:36:19.4 & 350.0 & 302.7 & 20.63$\pm$0.01 & 1.55$\pm$0.02 &  1085$\pm$ 45 &  ... &   ... &  1085$\pm$ 45 \nl
  282 & 12:43:20.43 & 11:30:26.8 & 329.5 & 240.4 & 20.75$\pm$0.02 & 1.38$\pm$0.02 &   883$\pm$ 94 &  ... &   ... &   883$\pm$ 94 \nl
  634 & 12:43:20.58 & 11:40:27.3 & 522.2 & 326.8 & 21.43$\pm$0.03 & 1.74$\pm$0.03 &  1009$\pm$ 62 &  ... &   ... &  1009$\pm$ 62 \nl
  382 & 12:43:21.62 & 11:35:50.1 & 313.2 & 300.7 & 20.98$\pm$0.02 & 1.48$\pm$0.02 &  1203$\pm$ 87 &  ... &   ... &  1203$\pm$ 87 \nl
  279 & 12:43:22.03 & 11:29:24.6 & 346.0 & 229.4 & 20.74$\pm$0.02 & 1.71$\pm$0.02 &  1033$\pm$ 43 &  ... &   ... &  1033$\pm$ 43 \nl
  120 & 12:43:22.62 & 11:29:45.5 & 326.0 & 231.2 & 20.06$\pm$0.06 & 1.63$\pm$0.06 &  1389$\pm$ 50 &  ... &   ... &  1389$\pm$ 50 \nl
  178 & 12:43:23.56 & 11:32:20.3 & 245.6 & 258.4 & 20.35$\pm$0.01 & 1.31$\pm$0.02 &  1212$\pm$ 42 &  ... &   ... &  1212$\pm$ 42 \nl
  245 & 12:43:25.01 & 11:31:17.7 & 246.1 & 242.9 & 20.63$\pm$0.01 & 1.34$\pm$0.02 &  1195$\pm$ 47 &  ... &   ... &  1195$\pm$ 47 \nl
  363 & 12:43:25.73 & 11:37:38.3 & 340.0 & 322.0 & 20.95$\pm$0.02 & 1.55$\pm$0.02 &   871$\pm$ 36 &  ... &   ... &   871$\pm$ 36 \nl
  308 & 12:43:25.94 & 11:30:18.3 & 267.5 & 230.2 & 20.82$\pm$0.01 & 1.90$\pm$0.02 &  1300$\pm$ 47 &  ... &   ... &  1300$\pm$ 47 \nl
  231 & 12:43:27.56 & 11:38:06.6 & 348.0 & 328.3 & 20.60$\pm$0.01 & 1.64$\pm$0.02 &  1431$\pm$104 &  ... &   ... &  1431$\pm$104 \nl
  394 & 12:43:29.22 & 11:31:26.9 & 187.9 & 236.9 & 21.00$\pm$0.01 & 1.49$\pm$0.02 &  1200$\pm$ 37 &  ... &   ... &  1200$\pm$ 37 \nl
  288 & 12:43:29.36 & 11:37:32.4 & 305.1 & 329.2 & 20.76$\pm$0.01 & 1.35$\pm$0.02 &  1245$\pm$ 39 &  ... &   ... &  1245$\pm$ 39 \nl
  205 & 12:43:29.48 & 11:34:53.8 & 185.6 & 304.0 & 20.50$\pm$0.05 & 1.79$\pm$0.09 &  1268$\pm$ 64 &  ... &   ... &  1268$\pm$ 64 \nl
  193 & 12:43:29.49 & 11:30:01.2 & 242.7 & 219.2 & 20.44$\pm$0.01 & 1.76$\pm$0.02 &  1076$\pm$ 49 &  ... &   ... &  1076$\pm$ 49 \nl
  248 & 12:43:30.08 & 11:31:33.2 & 174.0 & 236.4 & 20.64$\pm$0.02 & 1.66$\pm$0.02 &  1214$\pm$ 34 &  ... &   ... &  1214$\pm$ 34 \nl
  628 & 12:43:30.43 & 11:30:16.7 & 222.0 & 219.0 & 21.42$\pm$0.01 & 1.55$\pm$0.02 &   851$\pm$ 92 &  ... &   ... &   851$\pm$ 92 \nl
  404 & 12:43:31.38 & 11:33:02.6 & 126.2 & 266.8 & 21.02$\pm$0.04 & 1.64$\pm$0.05 &  1408$\pm$ 25 &  ... &   ... &  1408$\pm$ 25 \nl
  605 & 12:43:31.67 & 11:36:51.6 & 252.8 & 331.1 & 21.38$\pm$0.02 & 1.75$\pm$0.03 &  1310$\pm$ 34 &  ... &   ... &  1310$\pm$ 34 \nl
  269 & 12:43:32.55 & 11:30:32.3 & 190.9 & 214.7 & 20.71$\pm$0.01 & 1.63$\pm$0.02 &   929$\pm$ 19 &  ... &   ... &   929$\pm$ 19 \nl
  146 & 12:43:32.80 & 11:38:43.8 & 349.9 & 342.4 & 20.23$\pm$0.01 & 1.77$\pm$0.02 &  1587$\pm$ 28 &  ... &   ... &  1587$\pm$ 28 \nl
  654 & 12:43:33.28 & 11:38:54.9 & 358.5 & 344.0 & 21.45$\pm$0.02 & 1.52$\pm$0.02 &  1302$\pm$ 53 &  ... &   ... &  1302$\pm$ 53 \nl
  115 & 12:43:33.39 & 11:30:20.1 & 194.7 & 209.6 & 20.04$\pm$0.01 & 1.52$\pm$0.02 &  1046$\pm$ 29 &  ... &   ... &  1046$\pm$ 29 \nl
  153 & 12:43:33.43 & 11:28:39.1 & 286.4 & 199.5 & 20.23$\pm$0.01 & 1.36$\pm$0.01 &   885$\pm$ 34 &  ... &   ... &   885$\pm$ 34 \nl
  357 & 12:43:33.95 & 11:31:02.1 & 154.7 & 214.7 & 20.93$\pm$0.02 & 1.33$\pm$0.02 &  1264$\pm$106 &  ... &   ... &  1264$\pm$106 \nl
  362 & 12:43:34.57 & 11:32:27.6 &  89.6 & 242.0 & 20.95$\pm$0.02 & 1.40$\pm$0.03 &  1340$\pm$ 41 &  ... &   ... &  1340$\pm$ 41 \nl
  155 & 12:43:34.61 & 11:36:10.3 & 196.6 & 336.4 & 20.25$\pm$0.03 & 1.28$\pm$0.03 &  1586$\pm$ 37 &  ... &   ... &  1586$\pm$ 37 \nl
  201 & 12:43:34.76 & 11:31:30.0 & 125.3 & 217.5 & 20.48$\pm$0.01 & 1.39$\pm$0.02 &   649$\pm$ 52 &  ... &   ... &   649$\pm$ 52 \nl
  156 & 12:43:34.85 & 11:27:05.4 & 371.2 & 191.6 & 20.26$\pm$0.02 & 1.61$\pm$0.02 &   914$\pm$ 93 &  ... &   ... &   914$\pm$ 93 \nl
  194 & 12:43:35.12 & 11:28:45.3 & 273.1 & 195.0 & 20.45$\pm$0.01 & 1.65$\pm$0.02 &   853$\pm$107 &  ... &   ... &   853$\pm$107 \nl
  230 & 12:43:35.17 & 11:36:09.4 & 192.6 & 338.5 & 20.60$\pm$0.02 & 1.65$\pm$0.03 &  1480$\pm$ 21 &   664$\pm$ 56 &  1574 &  1480$\pm$ 21 \nl
  385 & 12:43:35.63 & 11:33:03.2 &  63.9 & 264.2 & 20.99$\pm$0.04 & 1.47$\pm$0.04 &  1157$\pm$ 57 &  ... &   ... &  1157$\pm$ 57 \nl
  119 & 12:43:36.07 & 11:31:19.6 & 123.7 & 207.4 & 20.07$\pm$0.02 & 1.91$\pm$0.03 &   944$\pm$ 37 &  ... &   ... &   944$\pm$ 37 \nl
  180 & 12:43:36.36 & 11:27:27.5 & 345.5 & 188.7 & 20.36$\pm$0.02 & 1.74$\pm$0.02 &   765$\pm$ 33 &  ... &   ... &   765$\pm$ 33 \nl
  124 & 12:43:36.53 & 11:39:53.0 & 406.0 & 352.7 & 20.08$\pm$0.01 & 1.65$\pm$0.01 &  1333$\pm$ 56 &  ... &   ... &  1333$\pm$ 56 \nl
  373 & 12:43:37.03 & 11:33:41.6 &  53.6 & 306.4 & 20.97$\pm$0.03 & 1.48$\pm$0.04 &  1014$\pm$ 63 &  ... &   ... &  1014$\pm$ 63 \nl
 9976 & 12:43:37.20 & 11:33:05.8 &  40.7 & 264.5 & 21.42$\pm$0.12 & 1.99$\pm$0.14 &  ... &   627$\pm$ 66 &  1443 &   627$\pm$ 66 \nl
  171 & 12:43:37.26 & 11:31:16.6 & 119.4 & 199.3 & 20.33$\pm$0.02 & 1.41$\pm$0.03 &  1028$\pm$ 89 &  ... &   ... &  1028$\pm$ 89 \nl
 9873 & 12:43:37.51 & 11:33:03.3 &  36.6 & 260.1 & 19.79$\pm$0.05 & 1.65$\pm$0.05 &  ... &  1101$\pm$ 60 &  1384 &  1101$\pm$ 60 \nl
  543 & 12:43:37.57 & 11:32:45.6 &  42.6 & 235.7 & 21.30$\pm$0.04 & 1.65$\pm$0.05 &  1144$\pm$ 48 &  ... &   ... &  1144$\pm$ 48 \nl
  264 & 12:43:37.74 & 11:37:03.1 & 235.2 & 351.9 & 20.69$\pm$0.02 & 1.81$\pm$0.03 &  1220$\pm$ 29 &  ... &   ... &  1220$\pm$ 29 \nl
  353 & 12:43:38.31 & 11:32:10.0 &  64.2 & 202.1 & 20.93$\pm$0.03 & 2.07$\pm$0.04 &  ... &  1312$\pm$ 39 &  1298 &  1312$\pm$ 39 \nl
  366 & 12:43:38.46 & 11:36:47.8 & 218.8 & 354.1 & 20.96$\pm$0.02 & 1.57$\pm$0.03 &   960$\pm$ 40 &  ... &   ... &   960$\pm$ 40 \nl
  143 & 12:43:38.51 & 11:30:27.3 & 163.3 & 187.4 & 20.20$\pm$0.05 & 1.36$\pm$0.06 &   916$\pm$ 50 &  ... &   ... &   916$\pm$ 50 \nl
  522 & 12:43:38.54 & 11:28:52.4 & 257.4 & 184.6 & 21.27$\pm$0.02 & 1.96$\pm$0.03 &   991$\pm$ 49 &  ... &   ... &   991$\pm$ 49 \nl
 9880 & 12:43:38.56 & 11:32:39.4 &  36.6 & 214.4 & 20.19$\pm$0.07 & 1.46$\pm$0.07 &  ... &  1113$\pm$ 48 &  1252 &  1113$\pm$ 48 \nl
 9948 & 12:43:38.81 & 11:33:46.6 &  40.6 & 335.2 & 21.20$\pm$0.07 & 2.00$\pm$0.08 &  ... &  1446$\pm$ 46 &  1182 &  1446$\pm$ 46 \nl
 9955 & 12:43:38.89 & 11:32:40.9 &  32.6 & 208.7 & 21.27$\pm$0.10 & 1.64$\pm$0.11 &  ... &   858$\pm$ 72 &  1211 &   858$\pm$ 72 \nl
   51 & 12:43:38.90 & 11:34:35.2 &  86.8 & 349.5 & 19.24$\pm$0.03 & 1.77$\pm$0.04 &   732$\pm$ 55 &  ... &   ... &   732$\pm$ 55 \nl
  563 & 12:43:39.04 & 11:31:38.9 &  91.4 & 188.4 & 21.32$\pm$0.03 & 1.41$\pm$0.03 &  1083$\pm$ 62 &  ... &   ... &  1083$\pm$ 62 \nl
  219 & 12:43:39.25 & 11:37:28.3 & 258.3 & 357.6 & 20.56$\pm$0.01 & 1.61$\pm$0.02 &  1143$\pm$ 50 &  ... &   ... &  1143$\pm$ 50 \nl
 9979 & 12:43:39.55 & 11:34:52.1 & 102.3 & 356.4 & 21.46$\pm$0.06 & 1.17$\pm$0.06 &  ... &  1291$\pm$ 40 &  1098 &  1291$\pm$ 40 \nl
  570 & 12:43:39.68 & 11:30:09.4 & 179.8 & 181.2 & 21.34$\pm$0.02 & 1.59$\pm$0.02 &   887$\pm$ 54 &  ... &   ... &   887$\pm$ 54 \nl
 9893 & 12:43:39.99 & 11:31:58.6 &  70.8 & 179.6 & 20.47$\pm$0.04 & 1.77$\pm$0.04 &  ... &   897$\pm$ 49 &  1145 &   897$\pm$ 49 \nl
 9916 & 12:43:40.35 & 11:33:49.3 &  39.9 &   8.0 & 20.84$\pm$0.06 & 1.34$\pm$0.06 &  1375$\pm$ 61 &  ... &   ... &  1375$\pm$ 61 \nl
  349 & 12:43:40.36 & 11:35:48.1 & 158.2 &   2.0 & 20.92$\pm$0.06 & 1.53$\pm$0.06 &  1134$\pm$ 50 &  1114$\pm$ 47 &  1011 &  1123$\pm$ 34 \nl
  439 & 12:43:40.42 & 11:31:10.1 & 119.4 & 176.7 & 21.11$\pm$0.03 & 1.37$\pm$0.04 &  1167$\pm$155 &  ... &   ... &  1167$\pm$155 \nl
  565 & 12:43:40.46 & 11:28:28.4 & 280.7 & 178.4 & 21.33$\pm$0.03 & 1.94$\pm$0.05 &   940$\pm$ 63 &  ... &   ... &   940$\pm$ 63 \nl
  443 & 12:43:40.47 & 11:31:33.6 &  96.0 & 175.6 & 21.11$\pm$0.03 & 1.42$\pm$0.04 &  ... &  1334$\pm$ 49 &  1126 &  1334$\pm$ 49 \nl
 9918 & 12:43:40.68 & 11:33:51.3 &  42.8 &  13.9 & 20.87$\pm$0.06 & 1.77$\pm$0.07 &  ... &   991$\pm$ 44 &  1037 &   991$\pm$ 44 \nl
  149 & 12:43:41.02 & 11:32:09.3 &  62.1 & 165.6 & 20.23$\pm$0.03 & 1.64$\pm$0.03 &   953$\pm$ 41 &   925$\pm$ 49 &  1063 &   942$\pm$ 31 \nl
   96 & 12:43:41.92 & 11:32:20.2 &  56.9 & 149.8 & 19.95$\pm$0.03 & 1.67$\pm$0.03 &  ... &  1028$\pm$ 44 &   975 &  1028$\pm$ 44 \nl
  218 & 12:43:42.08 & 11:28:53.1 & 257.8 & 173.0 & 20.55$\pm$0.01 & 1.77$\pm$0.02 &   848$\pm$ 47 &  ... &   ... &   848$\pm$ 47 \nl
  305 & 12:43:42.13 & 11:36:55.4 & 227.4 &   7.9 & 20.81$\pm$0.02 & 1.28$\pm$0.02 &  1387$\pm$ 55 &  ... &   ... &  1387$\pm$ 55 \nl
  190 & 12:43:42.54 & 11:26:55.4 & 375.5 & 174.1 & 20.41$\pm$0.02 & 1.57$\pm$0.02 &  1246$\pm$ 37 &  ... &   ... &  1246$\pm$ 37 \nl
  160 & 12:43:42.57 & 11:32:01.5 &  77.9 & 150.6 & 20.28$\pm$0.03 & 1.61$\pm$0.03 &  1359$\pm$ 19 &  1361$\pm$ 44 &   899 &  1359$\pm$ 17 \nl
  383 & 12:43:42.65 & 11:30:59.2 & 135.9 & 163.1 & 21.00$\pm$0.02 & 1.86$\pm$0.03 &   616$\pm$ 59 &  ... &   ... &   616$\pm$ 59 \nl
  141 & 12:43:43.05 & 11:32:20.0 &  67.0 & 137.5 & 20.19$\pm$0.03 & 2.18$\pm$0.04 &  ... &  1179$\pm$ 48 &   806 &  1179$\pm$ 48 \nl
  811 & 12:43:43.47 & 11:35:09.8 & 130.3 &  23.1 & 21.62$\pm$0.03 & 1.37$\pm$0.04 &  ... &   990$\pm$108 &   640 &   990$\pm$108 \nl
  529 & 12:43:43.60 & 11:32:54.7 &  55.3 & 105.5 & 21.27$\pm$0.03 & 1.53$\pm$0.04 &   856$\pm$ 43 &  ... &   ... &   856$\pm$ 43 \nl
  216 & 12:43:43.74 & 11:31:44.7 & 101.2 & 146.8 & 20.53$\pm$0.02 & 1.62$\pm$0.03 &  ... &   934$\pm$ 46 &   740 &   934$\pm$ 46 \nl
  140 & 12:43:43.76 & 11:32:56.8 &  57.0 & 102.9 & 20.18$\pm$0.04 & 1.68$\pm$0.04 &   825$\pm$ 32 &  ... &   ... &   825$\pm$ 32 \nl
  386 & 12:43:44.13 & 11:35:20.1 & 143.6 &  25.0 & 20.99$\pm$0.03 & 1.33$\pm$0.04 &   793$\pm$ 65 &  ... &   ... &   793$\pm$ 65 \nl
  167 & 12:43:44.81 & 11:30:56.6 & 150.5 & 151.7 & 20.32$\pm$0.01 & 1.55$\pm$0.02 &   724$\pm$ 66 &  ... &   ... &   724$\pm$ 66 \nl
  101 & 12:43:44.83 & 11:32:54.3 &  72.9 & 102.1 & 19.99$\pm$0.04 & 1.88$\pm$0.05 &  ... &   585$\pm$ 41 &   606 &   585$\pm$ 41 \nl
  187 & 12:43:45.24 & 11:33:18.3 &  77.7 &  83.6 & 20.38$\pm$0.05 & 1.46$\pm$0.06 &  1051$\pm$ 44 &  1065$\pm$ 52 &   558 &  1057$\pm$ 33 \nl
  359 & 12:43:45.79 & 11:39:37.6 & 396.6 &  12.3 & 20.94$\pm$0.01 & 1.31$\pm$0.02 &  1046$\pm$122 &  ... &   ... &  1046$\pm$122 \nl
  326 & 12:43:46.18 & 11:32:20.3 & 103.5 & 118.3 & 20.86$\pm$0.03 & 1.77$\pm$0.05 &   828$\pm$ 73 &   825$\pm$ 58 &   517 &   826$\pm$ 45 \nl
  337 & 12:43:46.24 & 11:33:34.4 &  95.2 &  74.9 & 20.90$\pm$0.03 & 2.07$\pm$0.03 &  ... &   649$\pm$ 45 &   502 &   649$\pm$ 45 \nl
  547 & 12:43:46.39 & 11:37:51.5 & 296.6 &  18.4 & 21.30$\pm$0.01 & 1.22$\pm$0.02 &  1477$\pm$ 69 &  ... &   ... &  1477$\pm$ 69 \nl
  452 & 12:43:47.35 & 11:31:56.1 & 130.7 & 124.0 & 21.13$\pm$0.02 & 1.51$\pm$0.03 &  ... &  1091$\pm$ 47 &   462 &  1091$\pm$ 47 \nl
  324 & 12:43:47.42 & 11:29:06.9 & 265.7 & 155.6 & 20.85$\pm$0.02 & 1.57$\pm$0.03 &  1183$\pm$ 67 &  ... &   ... &  1183$\pm$ 67 \nl
  296 & 12:43:47.55 & 11:32:21.9 & 120.9 & 113.1 & 20.78$\pm$0.02 & 1.61$\pm$0.03 &   886$\pm$ 57 &  ... &   ... &   886$\pm$ 57 \nl
  384 & 12:43:47.86 & 11:31:42.8 & 144.6 & 126.7 & 20.99$\pm$0.02 & 1.71$\pm$0.02 &  ... &   896$\pm$ 38 &   434 &   896$\pm$ 38 \nl
  144 & 12:43:47.93 & 11:35:40.0 & 190.0 &  37.8 & 20.21$\pm$0.01 & 1.49$\pm$0.02 &  ... &  1255$\pm$ 41 &   360 &  1255$\pm$ 41 \nl
   99 & 12:43:48.18 & 11:32:35.7 & 125.2 & 105.6 & 19.98$\pm$0.02 & 1.95$\pm$0.02 &   919$\pm$ 21 &  ... &   ... &   919$\pm$ 21 \nl
  320 & 12:43:49.85 & 11:32:18.6 & 153.7 & 109.3 & 20.84$\pm$0.02 & 1.44$\pm$0.03 &  1517$\pm$ 71 &  1506$\pm$ 68 &   329 &  1511$\pm$ 49 \nl
  189 & 12:43:50.05 & 11:30:58.8 & 197.3 & 131.3 & 20.40$\pm$0.01 & 1.44$\pm$0.02 &   931$\pm$ 34 &  ... &   ... &   931$\pm$ 34 \nl
 1081 & 12:43:50.25 & 11:32:11.3 & 161.6 & 111.0 & 21.89$\pm$0.03 & 1.72$\pm$0.05 &  ... &  1154$\pm$ 50 &   318 &  1154$\pm$ 50 \nl
  464 & 12:43:50.42 & 11:35:59.7 & 228.6 &  42.0 & 21.16$\pm$0.01 & 1.81$\pm$0.03 &  ... &  1077$\pm$ 45 &   251 &  1077$\pm$ 45 \nl
  676 & 12:43:50.50 & 11:33:04.9 & 154.5 &  91.7 & 21.49$\pm$0.03 & 1.49$\pm$0.04 &  ... &  1039$\pm$ 54 &   298 &  1039$\pm$ 54 \nl
  607 & 12:43:51.08 & 11:32:52.8 & 163.8 &  95.8 & 21.38$\pm$0.02 & 1.83$\pm$0.03 &  ... &  1285$\pm$ 48 &   277 &  1285$\pm$ 48 \nl
  133 & 12:43:51.37 & 11:31:31.7 & 193.8 & 120.2 & 20.12$\pm$0.01 & 1.40$\pm$0.02 &   738$\pm$ 59 &  ... &   ... &   738$\pm$ 59 \nl
  508 & 12:43:53.14 & 11:33:01.1 & 193.4 &  92.4 & 21.24$\pm$0.02 & 1.60$\pm$0.03 &   824$\pm$ 60 &  ... &   ... &   824$\pm$ 60 \nl
 1201 & 12:43:53.17 & 11:35:20.6 & 233.5 &  55.9 & 21.98$\pm$0.02 & 2.00$\pm$0.05 &  ... &  1245$\pm$ 47 &   158 &  1245$\pm$ 47 \nl
  623 & 12:43:53.34 & 11:31:36.1 & 217.2 & 115.4 & 21.41$\pm$0.02 & 1.33$\pm$0.03 &  1177$\pm$ 78 &  ... &   ... &  1177$\pm$ 78 \nl
  414 & 12:43:53.44 & 11:32:18.3 & 204.2 & 104.4 & 21.04$\pm$0.02 & 1.84$\pm$0.03 &  ... &  1285$\pm$ 37 &   183 &  1285$\pm$ 37 \nl
  280 & 12:43:53.72 & 11:32:17.3 & 208.4 & 104.4 & 20.74$\pm$0.02 & 1.83$\pm$0.02 &  1124$\pm$ 37 &  ... &   ... &  1124$\pm$ 37 \nl
  389 & 12:43:53.88 & 11:31:54.0 & 217.5 & 110.2 & 21.00$\pm$0.02 & 1.94$\pm$0.03 &  ... &   820$\pm$ 39 &   175 &   820$\pm$ 39 \nl
  142 & 12:43:54.18 & 11:27:39.1 & 390.6 & 147.6 & 20.18$\pm$0.02 & 1.58$\pm$0.02 &   922$\pm$ 34 &  ... &   ... &   922$\pm$ 34 \nl
  601 & 12:43:54.18 & 11:33:35.6 & 210.1 &  82.8 & 21.38$\pm$0.02 & 1.75$\pm$0.03 &  1108$\pm$ 40 &  ... &   ... &  1108$\pm$ 40 \nl
  746 & 12:43:54.50 & 11:32:42.2 & 214.9 &  97.2 & 21.57$\pm$0.02 & 1.81$\pm$0.04 &  ... &  1166$\pm$ 48 &   148 &  1166$\pm$ 48 \nl
  519 & 12:43:55.32 & 11:34:04.8 & 231.9 &  76.2 & 21.26$\pm$0.03 & 1.60$\pm$0.04 &  1141$\pm$ 88 &  ... &   ... &  1141$\pm$ 88 \nl
  317 & 12:43:55.46 & 11:32:12.0 & 234.4 & 104.1 & 20.83$\pm$0.01 & 1.29$\pm$0.02 &   785$\pm$ 29 &   800$\pm$ 62 &   124 &   788$\pm$ 26 \nl
  352 & 12:43:55.86 & 11:32:16.1 & 239.2 & 102.8 & 20.92$\pm$0.02 & 1.38$\pm$0.02 &  1005$\pm$ 50 &  ... &   ... &  1005$\pm$ 50 \nl
  681 & 12:43:56.17 & 11:32:11.9 & 244.5 & 103.5 & 21.50$\pm$0.02 & 1.78$\pm$0.03 &  ... &  1181$\pm$ 45 &    89 &  1181$\pm$ 45 \nl
  525 & 12:43:56.67 & 11:32:21.6 & 249.9 & 101.0 & 21.27$\pm$0.02 & 1.96$\pm$0.03 &   614$\pm$ 38 &   608$\pm$ 35 &    68 &   611$\pm$ 25 \nl
  274 & 12:43:57.82 & 11:31:12.1 & 287.1 & 114.0 & 20.73$\pm$0.02 & 1.62$\pm$0.03 &  1140$\pm$ 99 &  ... &   ... &  1140$\pm$ 99 \nl
  173 & 12:43:57.91 & 11:29:04.1 & 359.9 & 132.9 & 20.33$\pm$0.01 & 1.55$\pm$0.02 &   914$\pm$ 63 &  ... &   ... &   914$\pm$ 63 \nl
  159 & 12:43:59.97 & 11:28:23.5 & 409.8 & 134.1 & 20.28$\pm$0.02 & 1.56$\pm$0.02 &   877$\pm$ 37 &  ... &   ... &   877$\pm$ 37 \nl
  297 & 12:44:00.63 & 11:29:43.1 & 366.8 & 124.1 & 20.79$\pm$0.01 & 1.46$\pm$0.02 &  1152$\pm$ 90 &  ... &   ... &  1152$\pm$ 90 \nl
  632 & 12:44:08.94 & 11:27:55.1 & 529.4 & 126.3 & 21.42$\pm$0.07 & 1.30$\pm$0.08 &  1279$\pm$ 60 &  ... &   ... &  1279$\pm$ 60 \nl
\multicolumn{11}{c}{}\nl \multicolumn{11}{c}{\underbar{Foreground
Stars}}\nl \multicolumn{11}{c}{}\nl
  395 & 12:43:08.92 & 11:31:59.0 & 461.2 & 261.1 & 21.01$\pm$0.08 & 2.19$\pm$0.10 &    $-$1$\pm$ 54 &  ... &   ... &    $-$1$\pm$ 54 \nl
  495 & 12:43:12.26 & 11:33:45.1 & 408.2 & 274.9 & 21.22$\pm$0.02 & 1.62$\pm$0.03 &  $-$432$\pm$ 43 &  ... &   ... &  $-$432$\pm$ 43 \nl
 9883 & 12:43:12.34 & 11:35:55.1 & 438.0 & 292.1 & 20.32$\pm$0.09 & 3.20$\pm$0.10 &   229$\pm$ 60 &  ... &   ... &   229$\pm$ 60 \nl
  243 & 12:43:13.51 & 11:40:22.2 & 581.8 & 317.9 & 20.62$\pm$0.03 & 1.35$\pm$0.03 &   354$\pm$ 60 &  ... &   ... &   354$\pm$ 60 \nl
  344 & 12:43:22.92 & 11:38:04.9 & 386.8 & 319.6 & 20.90$\pm$0.02 & 1.13$\pm$0.02 &    13$\pm$ 39 &  ... &   ... &    13$\pm$ 39 \nl
  332 & 12:43:23.98 & 11:36:11.9 & 296.9 & 307.7 & 20.87$\pm$0.02 & 1.25$\pm$0.02 &   239$\pm$ 53 &  ... &   ... &   239$\pm$ 53 \nl
  271 & 12:43:33.79 & 11:38:45.2 & 347.1 & 344.8 & 20.71$\pm$0.02 & 1.55$\pm$0.02 &    $-$8$\pm$ 24 &  ... &   ... &    $-$8$\pm$ 24 \nl
  240 & 12:43:37.68 & 11:28:19.4 & 291.5 & 186.5 & 20.62$\pm$0.01 & 1.47$\pm$0.02 &    55$\pm$ 43 &  ... &   ... &    55$\pm$ 43 \nl
  128 & 12:43:45.90 & 11:30:53.6 & 161.2 & 147.2 & 20.09$\pm$0.01 & 1.36$\pm$0.02 &    11$\pm$ 37 &  ... &   ... &    11$\pm$ 37 \nl
  661 & 12:43:47.01 & 11:34:57.3 & 148.8 &  43.7 & 21.47$\pm$0.03 & 1.71$\pm$0.04 &  ... &   435$\pm$ 49 &   412 &   435$\pm$ 49 \nl
  461 & 12:43:48.48 & 11:33:29.6 & 126.4 &  80.9 & 21.15$\pm$0.03 & 1.58$\pm$0.03 &  $-$106$\pm$ 62 &  ... &   ... &  $-$106$\pm$ 62 \nl
  220 & 12:43:48.85 & 11:26:52.3 & 398.8 & 160.8 & 20.55$\pm$0.02 & 1.10$\pm$0.02 &   157$\pm$ 33 &  ... &   ... &   157$\pm$ 33 \nl
\multicolumn{11}{c}{}\nl \multicolumn{11}{c}{\underbar{Background
Galaxies}}\nl \multicolumn{11}{c}{}\nl
  444\tablenotemark{a} & 12:43:18.05 & 11:28:30.1 & 425.9 & 229.0 & 21.11$\pm$0.02 & 1.16$\pm$0.02 &   976$\pm$ 42 &  ... &   ... &   976$\pm$ 42 \nl
 9930 & 12:43:27.75 & 11:32:17.2 & 186.6 & 253.7 & 21.01$\pm$0.06 & 1.83$\pm$0.07 &  1471$\pm$134 &  ... &   ... &  1471$\pm$134 \nl
  176 & 12:43:33.62 & 11:29:43.3 & 225.8 & 204.3 & 20.34$\pm$0.06 & 1.68$\pm$0.07 &   826$\pm$ 43 &  ... &   ... &   826$\pm$ 43 \nl
99213 & 12:43:35.56 & 11:28:11.4 & 305.4 & 192.2 & 18.04$\pm$0.00 & 1.30$\pm$0.00 &  3839$\pm$ 66 &  ... &   ... &  3839$\pm$ 66 \nl
 9888 & 12:43:43.17 & 11:26:41.4 & 390.5 & 173.0 & 20.45$\pm$0.09 & 2.40$\pm$0.10 &  1074$\pm$ 66 &  ... &   ... &  1074$\pm$ 66 \nl
99214\tablenotemark{b} & 12:43:51.08 & 11:28:01.7 & 348.4 & 152.0 &  14.41$\pm$0.00 & 2.07$\pm$0.00  &  1027$\pm$ 53 &  ... &   ... &  1027$\pm$
53 \nl

\enddata
\tablenotetext{a~}{VCC 1963 in the Virgo Cluster catalog of \citet{bin85}}
\tablenotetext{b~}{VCC 1982 in the Virgo Cluster catalog of \citet{bin85}}
\end{deluxetable}
\clearpage
\begin{figure}
\plotone{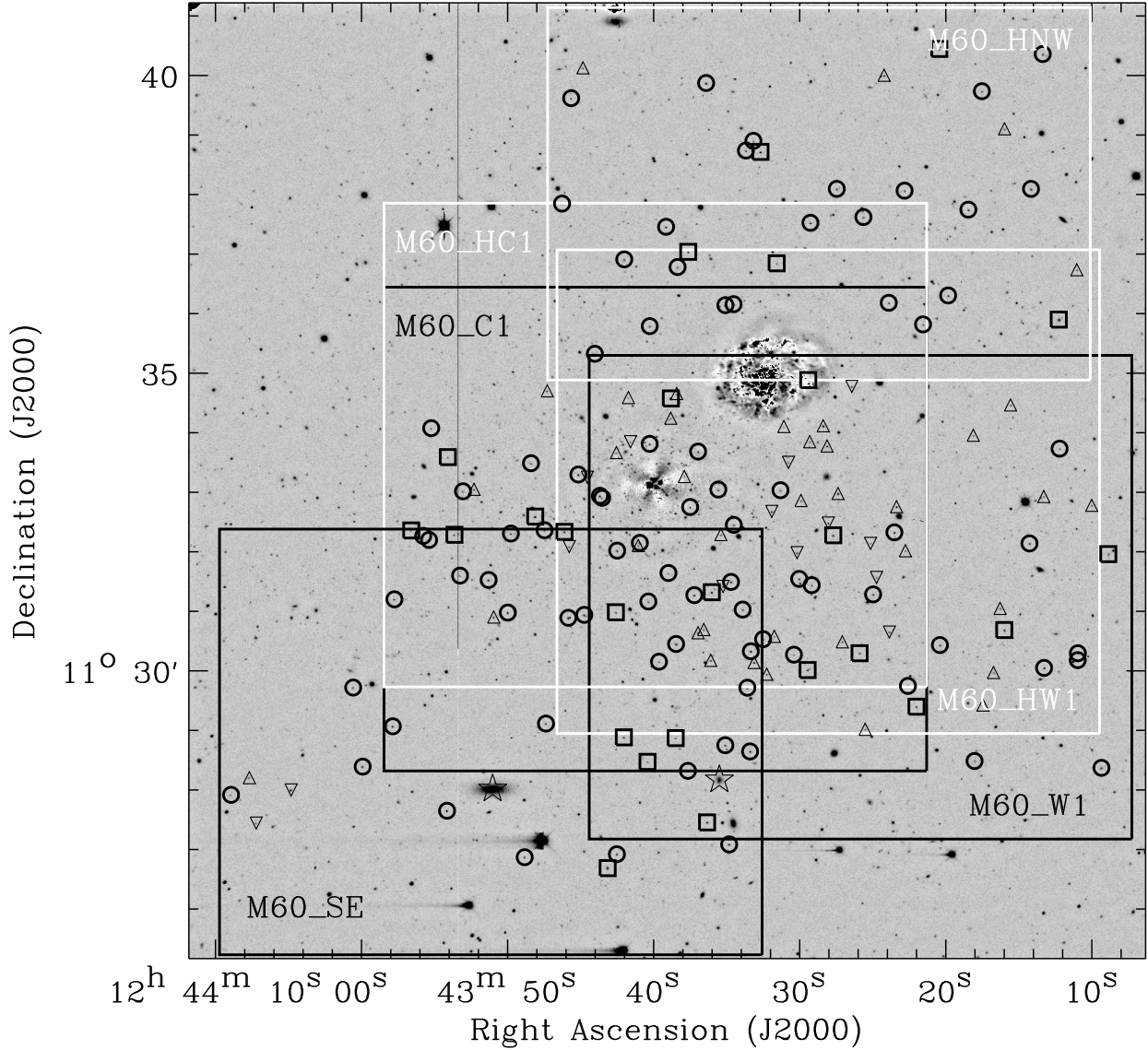}
\caption{
Positions of the observed masks (large squares) overlaid on the grayscale map of
the Washington $T_1$ image ($16.4\arcmin \times 16.4\arcmin$)
with the spectroscopic sample of globular cluster
 candidates.
The masks drawn with black lines and white lines were observed in 2002
and 2003 CFHT run, respectively.
The galaxy light of M60 was subtracted from the original image using IRAF/ELLIPSE fitting.
Open circles and open squares represent, respectively, the blue and the red globular clusters.
Triangles and upside-down triangles denote, respectively,
the positions of observed blue and red globular cluster candidates  of which spectra could not be extracted.
Two starlets represent background galaxies (ID: 99213 \& 99214).
\label{fig01}}
\end{figure}

\begin{figure}
\plotone{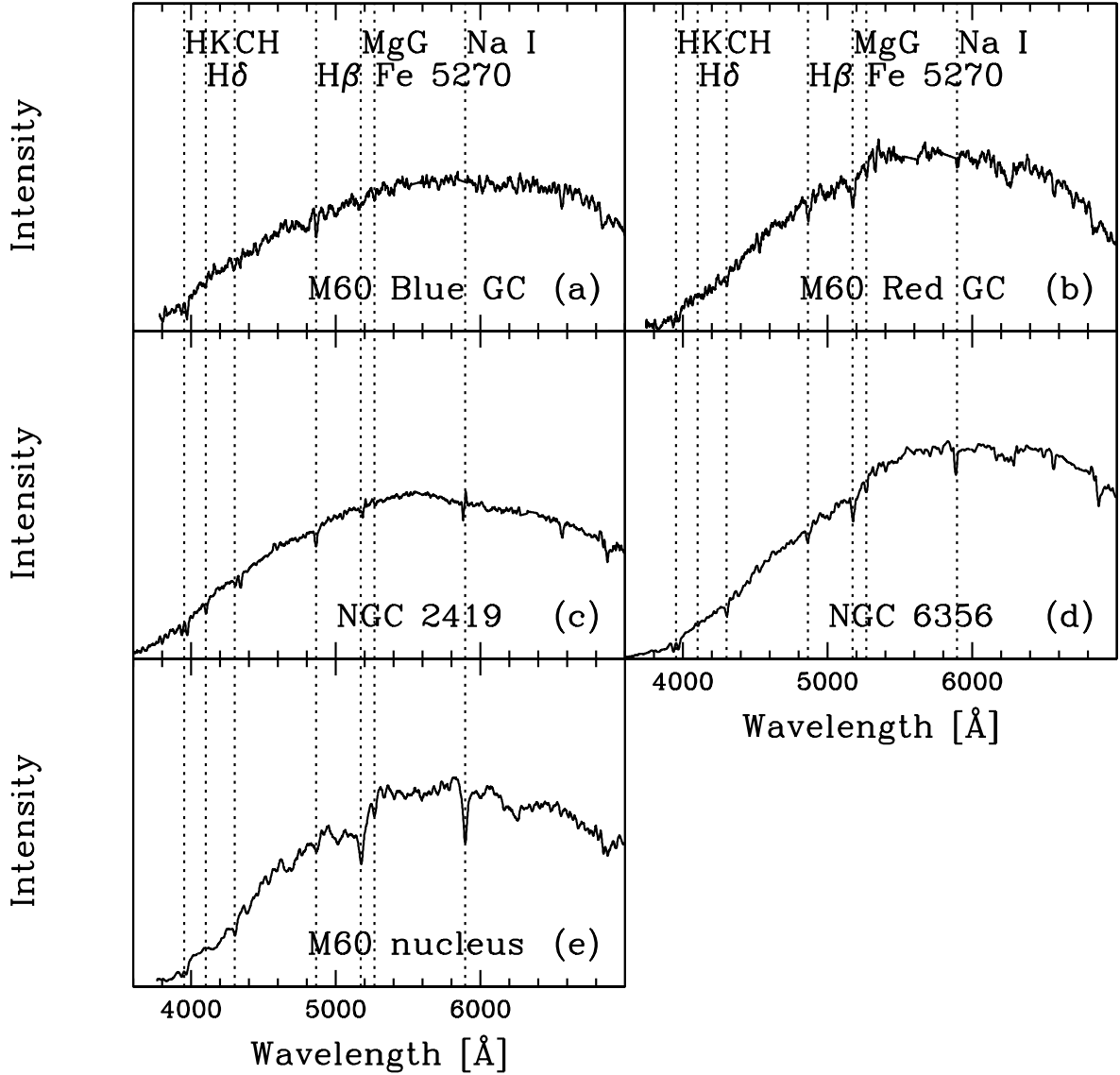}
\caption{
Example spectra of
(a) a blue globular cluster in M60 (ID: 151) with $T_1=20.22$ and $C-T_1=1.32$;
(b) a red  globular cluster in M60 (ID: 146) with $T_1=20.23$ and $C-T_1=1.77$;
(c) NGC 2419, a metal-poor Galactic globular cluster with [Fe/H]=$-$2.12 dex;
(d) NGC 6356, a metal-rich Galactic globular cluster with [Fe/H]=$-$0.50 dex; and
(e) M60 nucleus.
All spectra were smoothed and plotted in the rest frame.
\label{fig02}} 
\end{figure}

\begin{figure}
\plotone{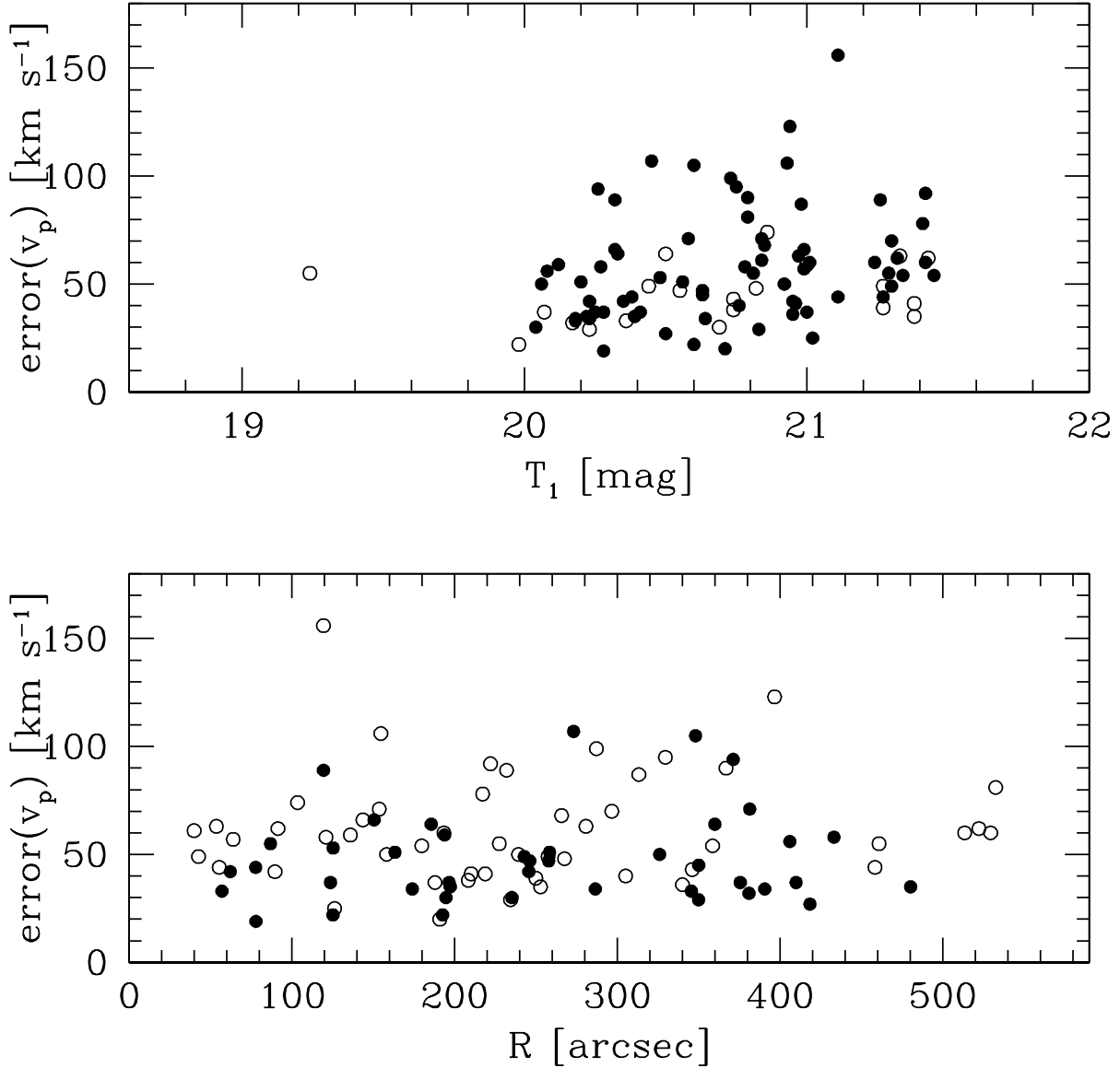}
\caption{{\it Upper panel}: Uncertainties of the measured radial velocities versus $T_1$ magnitudes
for 110 spectroscopic targets in this study.
The blue ($(C-T_1 ) <1.7$) and red ($(C-T_1 ) >1.7$) objects are represented by filled and open symbols, respectively.
{\it Lower panel}: Uncertainties of the measured radial velocities versus
galactocentric distances for 110 spectroscopic targets in this study.
The bright objects ($T_1<20.7$) and faint objects ($T_1\ge20.7$)
are represented by filled and open symbols, respectively.
\label{fig03}} 
\end{figure}

\begin{figure}
\plotone{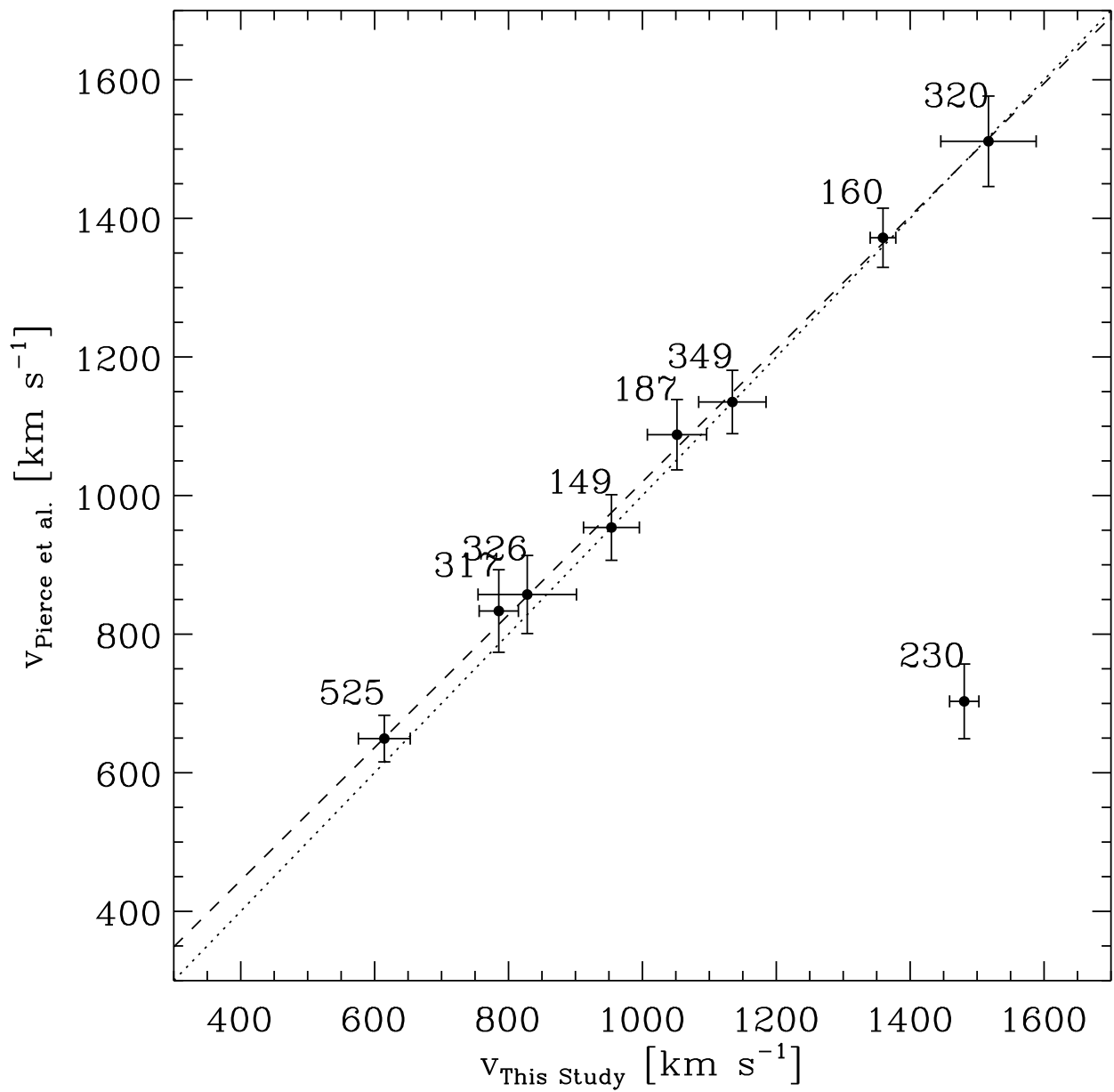} \caption{Comparison of radial velocities for M60 globular cluster candidates
measured in this study and in \citet{pie06}. The dashed line indicates the least-squares fit,
and the dotted line denotes the one-to-one relation.
The numbers represent the identifications of the globular cluster candidates in this study.
\label{fig04}} 
\end{figure}

\begin{figure}
\plotone{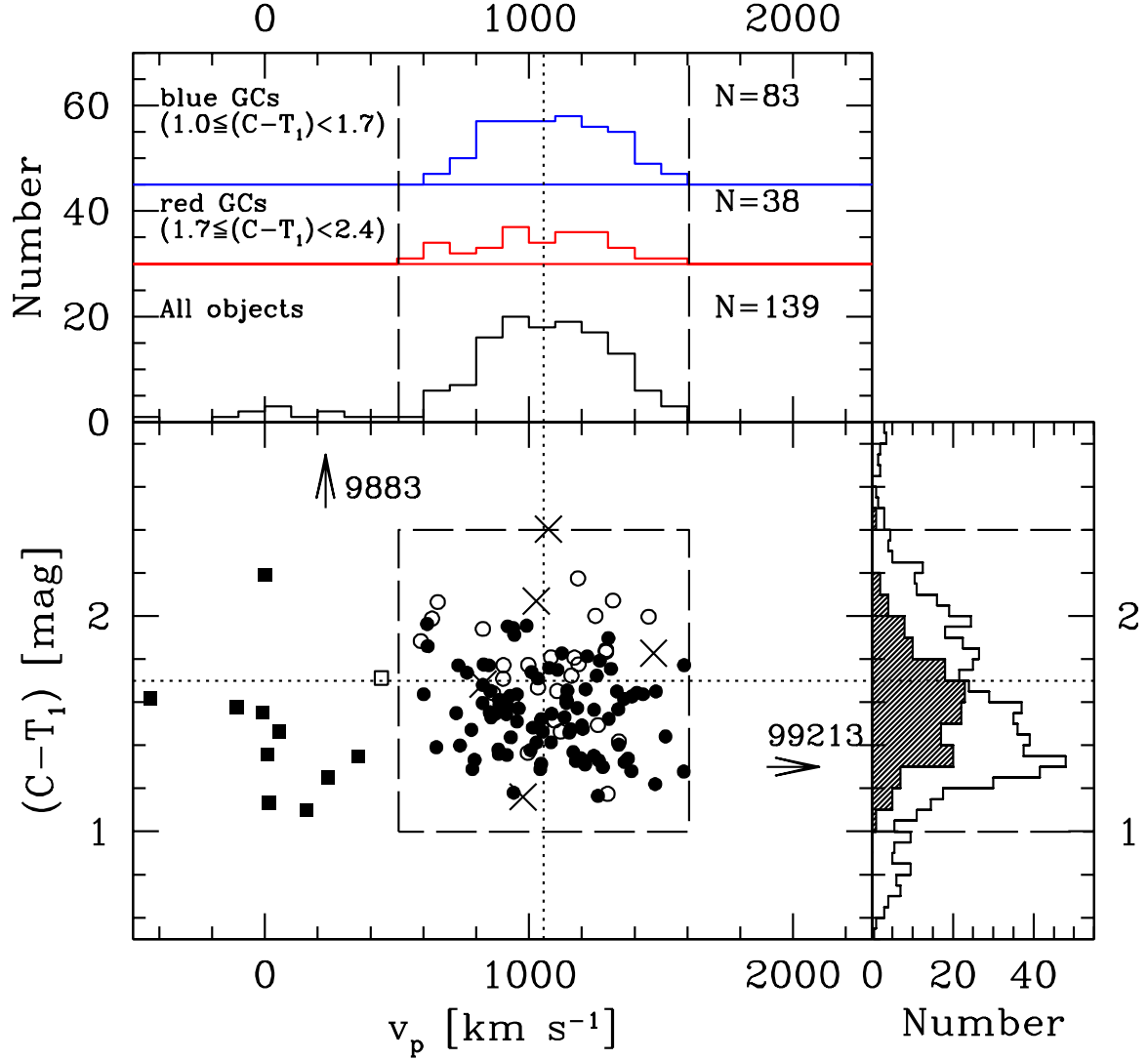} \caption{$(C-T_1)$ colors versus radial
velocities for 110 objects (filled symbols) derived in this
study and 29 objects (open symbols) measured only by \citet{pie06}.
The dashed line box indicates the selection criteria
of radial velocities and the $(C-T_1)$ colors for genuine globular clusters in M60.
Crosses in the dashed line box denote small galaxies, and filled squares and one open square represent the foreground stars.
The dotted vertical line indicates the velocity for the M60 nucleus, and the dotted
horizontal line denotes the color value which divides the globular clusters into
blue and red subsamples.
Two objects lying outside the plotted region are shown by arrows with their IDs.
{\it Upper panels:} The
radial velocity distributions for all objects, 38 red and 83 blue
genuine globular clusters in M60 are shown (from bottom to top).
{\it Lower right panel:} the color distribution for all objects is
shown by the shaded histogram, while that for the photometric sample of globular cluster
candidates  at 1$\arcmin<$R$<7\arcmin$ \citep{lee07}
is shown by the open histogram. \label{fig05}}
\end{figure}

\begin{figure}
\plotone{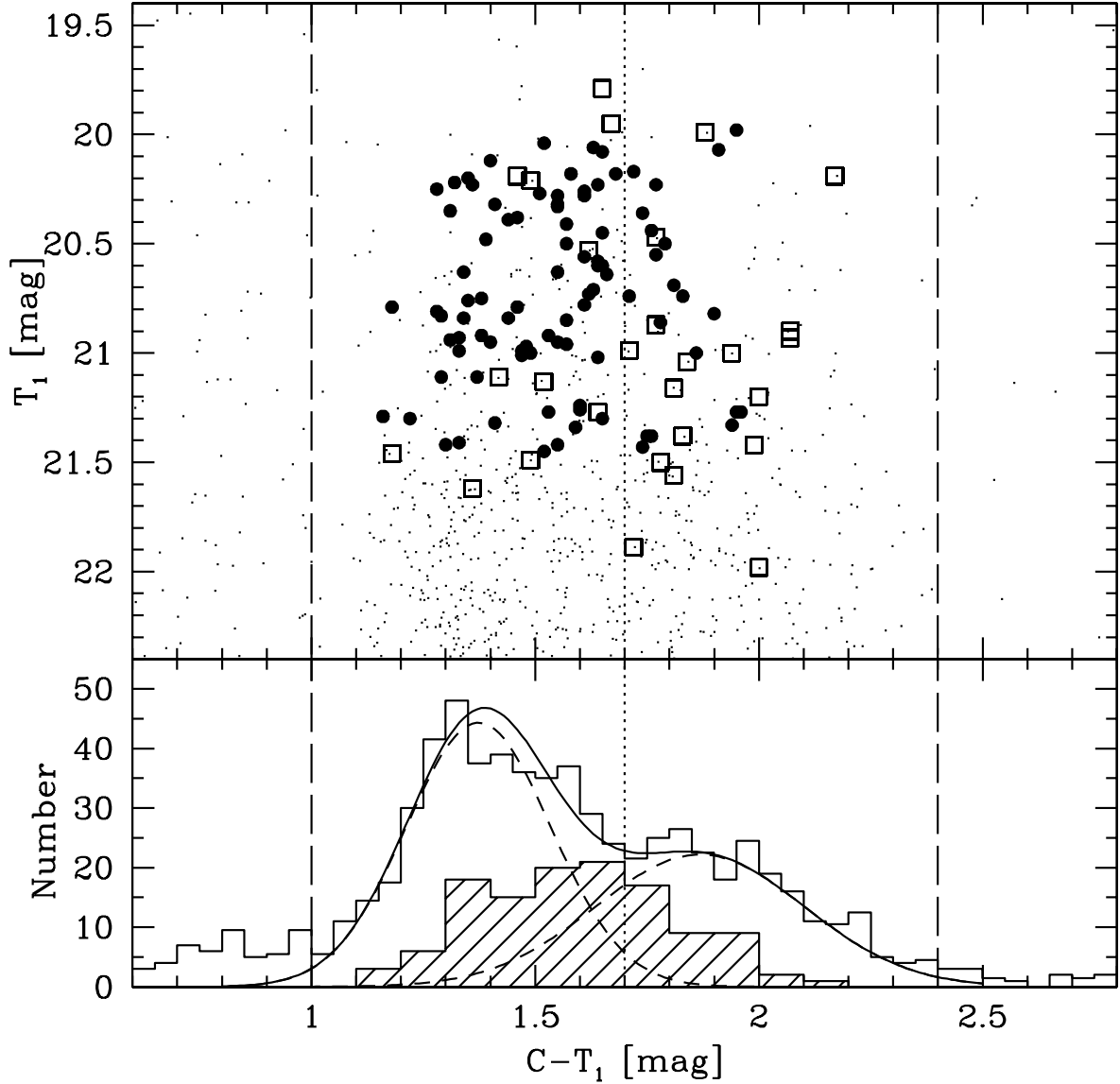}
\caption{{\it Upper panel}: Color magnitude diagram
for 121 genuine globular cluster in M60 from this study and \citet{pie06}. Filled circles indicate the
globular clusters measured in this study, and open squares are the globular clusters measured in
\citet{pie06}. Dots denote the photometric sample of M60 globular cluster candidates at 60$\arcsec<$R$<420\arcsec$ 
\citep{lee07}. 
The dotted vertical line denotes the color value that divides the globular clusters into
the blue and the red subsamples,
and the dashed vertical lines denote $(C-T_1)$ color boundary for M60 globular cluster candidates.
{\it Lower panel}: $(C-T_1)$ color distribution for 121 genuine globular clusters (hatched histogram)
compared with that for the photometric sample of globular cluster candidates (open histogram, \citealt{lee07}).
The thick solid line represents the double Gaussian fits to the photometric sample,
each component of which is plotted by the dashed lines.
\label{fig06}}
\end{figure}

\begin{figure}
\plotone{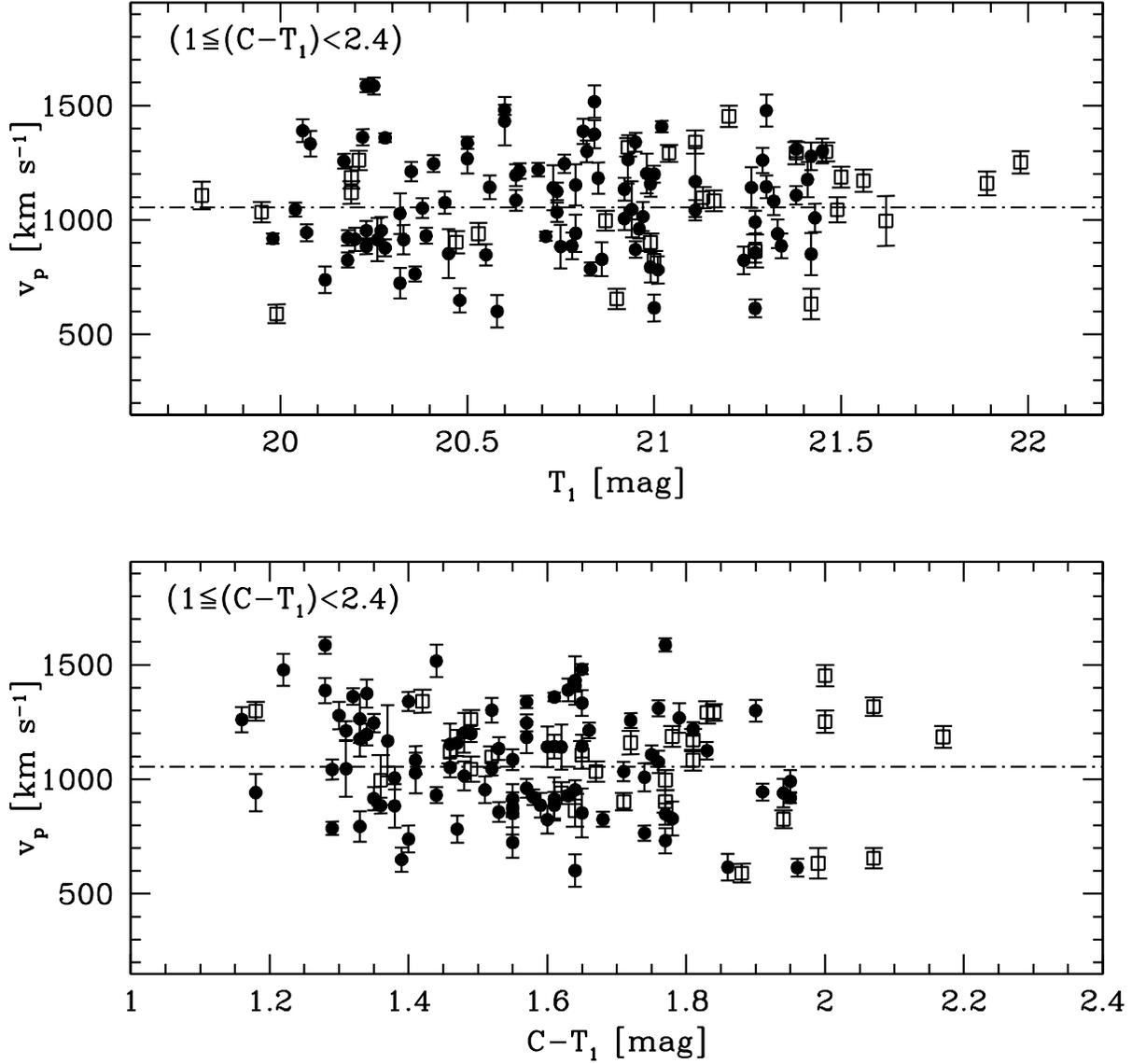}
\caption{Radial velocities with measured errors versus $T_1$ magnitudes (upper panel), and versus $C-T_1$ colors (lower panel)
for the genuine globular clusters in M60.
Filled circles  and open squares represent the globular clusters measured in this study
and in \citet{pie06}, respectively.
The dot-dashed horizontal line indicates the systemic velocity of M60.
\label{fig07}}
\end{figure}

\begin{figure}
\plotone{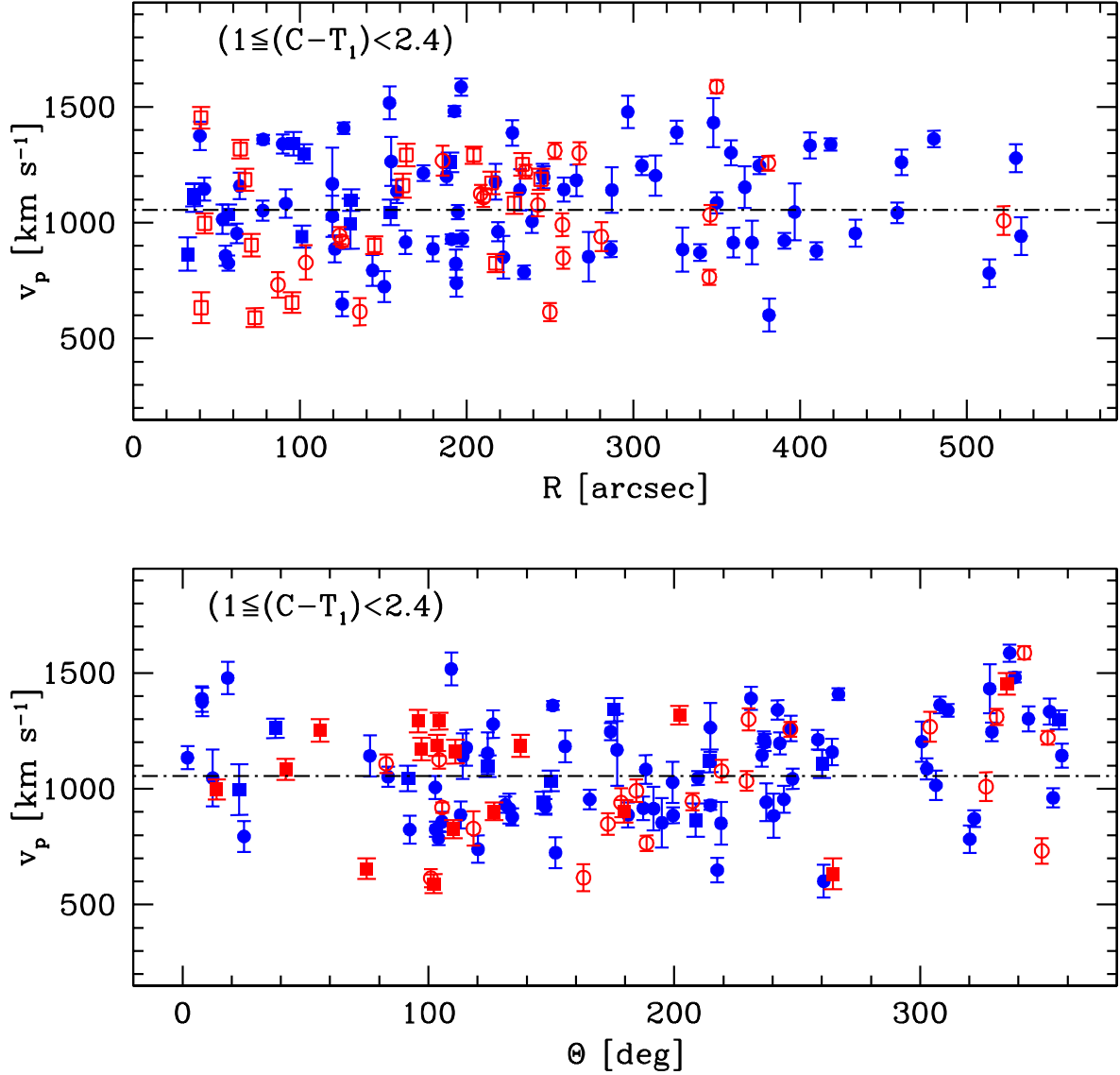}
\caption{Radial velocities with measured errors versus the projected galactocentric radii (upper panel),
and position angles (lower panel). The blue globular clusters and the red globular clusters
are represented, respectively, by the filled symbols
and the open symbols.
Circles and squares represent the globular clusters measured in this study
and \citet{pie06}, respectively.
The dot-dashed horizontal line indicates the systemic velocity of M60.
\label{fig08}}
\end{figure}

\clearpage


\begin{thebibliography}{}
\bibitem[Armandroff \& Zinn(1988)]{az88} Armandroff, T.~E.,
\& Zinn, R.\ 1988, \aj, 96, 92

\bibitem[Armosky et al.(1994)]{arm94} Armosky, B.~J., Sneden,
C., Langer, G.~E., \& Kraft, R.~P.\ 1994, \aj, 108, 1364

\bibitem[Ashman et al.(1994)]{ash94} Ashman, K.~M., Bird, 
C.~M., \& Zepf, S.~E.\ 1994, \aj, 108, 2348 


\bibitem[Beers et al.(1990a)]{beers90a} Beers, T. C., Flynn, K., \& Gebhardt, K. 1990a, \aj, 100, 32
\bibitem[Beers et al.(1990b)]{beers90b} Beers, T.~C., Kage,
J.~A., Preston, G.~W., \& Shectman, S.~A.\ 1990b, \aj, 100, 849

\bibitem[Binggeli et al.(1985)]{bin85} Binggeli, B., Sandage,
A., \& Tammann, G.~A.\ 1985, \aj, 90, 1681
\bibitem[Bridges et al.(2006)]{bri06}Bridges, T., et al. 2006, \mnras, 373, 157
\bibitem[Brodie \& Strader(2006)]{bro06} Brodie, J. P., \& Strader, J. 2006, \araa, 44, 193
\bibitem[Cohen \& McCarthy(1997)]{cm97} Cohen, J.~G., \& McCarthy, J.~K.\ 1997, \aj, 113, 1353

\bibitem[Cohen \& Ryzhov(1997)]{coh97} Cohen, J.G., \& Ryzhov, A. 1997, \apj, 486, 230
\bibitem[Cohen, Blakeslee \& C\^ot\'e (2003)]{coh03} Cohen, J. G., Blakeslee, J. P., C\^ot\'e, P. 2003, \aj, 592, 866
\bibitem[C\^ot\'e et al.(2001)]{cot01} C\^ot\'e, P., et al. 2001, \apj, 559, 828
\bibitem[C\^ot\'e et al.(2003)]{cot03} C\^ot\'e P., McLaughlin D. E., Cohen J. G., Blakeslee J. P. 2003, \apj, 591, 850
\bibitem[Couture et al.(1991)]{cou91} Couture, J., Harris, W. E., \& Allwright, J. W. B. 1991, \apj, 372, 97
\bibitem[de Bruyne et al.(2001)]{deb01} de Bruyne, V., Dejonghe, H.,
Pizzella, A., Bernardi, M., \& Zeilinger, W. W. 2001, \apj, 546, 903
\bibitem[Forbes et al.(2004)]{for04} Forbes D.A., et al. 2004, \mnras, 355, 608
\bibitem[Gonz\'alez(1993)]{gon93} Gonz\'alez, J. J., 1993, PhD thesis,
Univ. California, Santa Cruz
\bibitem[Harris(1996)]{har96} Harris, W. E. 1996, \aj, 112, 1487 (February 2003 version)
\bibitem[Harris et al.(1991)]{har91} Harris, W. E., Allwright, J. W. B., Pritchet, C. J., van den Bergh, S. 1991, \apjs, 76, 115
\bibitem[Harris \& Harris(2002)]{har02} Harris, W. E., \& Harris, G. L. H. 2002, \aj, 123, 3108

\bibitem[Hesser et al.(1986)]{hsm86} Hesser, J.~E., Shawl,
S.~J., \& Meyer, J.~E.\ 1986, \pasp, 98, 403

\bibitem[Hwang et al.(2007)]{hwa07}Hwang, H. S., et al. 2007, \apj, submitted (Paper II)
\bibitem[Kim et al.(2006)]{kim06} Kim, E., Kim, D.-W., Fabbiano, G.,
Lee, M. G., Park, H. S., Geisler, D., Dirsch, B. 2006, \apj, 647, 276
\bibitem[Kissler-Patig et al.(1998)]{kis98} Kissler-Patig, M., Brodie, J.P.,
Schroder, L.L., Forbes, D.A., Grillmair, C.J., \& Huchra, J.P. 1998, \aj, 115, 105
\bibitem[Kissler-Patig \& Gebhardt(1998)]{kis98b} Kissler-Patig, M., \&
Gebhardt, K.  1998, \aj, 116, 2237
\bibitem[Kissler-Patig et al.(1999)]{kis99} Kissler-Patig M., Grillmair C.J.,
Meylan G., Brodie, J.P., Minniti, D., \& Goudfrooij, P. 1999, \aj, 117, 1206
\bibitem[Koopman et al.(2001)]{koo01} Koopman, R.A., Kenney, J.D.P., \&
Young, J. 2001, \apjs, 135, 125
\bibitem[Kundu \& Whitmore(2001)]{kun01} Kundu, A., \& Whitmore, B.C. 2001,
\aj, 121, 2950
\bibitem[Larsen et al.(2001)]{lar01} Larsen, S.S., Brodie, J.P., Huchra, J.P.,
Forbes, D.A., \& Grillmair, C.J. 2001, \aj, 121, 2974
\bibitem[Lee(2003)]{lee03} Lee, M. G. 2003, J.  Korean  Astro. Soc., 36, 189
\bibitem[Lee et al.(2007)]{lee07} Lee, M. G., Park, H.S., Kim, E., Hwang, H. S., Kim, S. C. \& Geisler, D. 2007, \apj, submitted
\bibitem[Le Fevre et al.(1994)]{lef94} Le Fevre, O., Crampton, D., Felenbok, P.,
  \& Monnet, G. 1994, \aap, 282, 325
\bibitem[Mei \etal (2007)]{mei07} Mei, S. \etal 2007, \apj, 655, 144 
\bibitem[Mieske et al.(2006)]{mie06} Mieske, S., et al. 2006, \apj, 653, 193 

\bibitem[Minniti(1995a)]{min95a} Minniti, D.\ 1995a, \aaps, 113, 299
\bibitem[Minniti(1995b)]{min95b} Minniti, D.\ 1995b, \aap, 303, 468

\bibitem[Minniti et al.(1998)]{min98} Minniti, D., Kissler-Patig, M., Goudfrooij, P.,
\& Meylan, G.  1998, \aj, 115, 121
\bibitem[Neilsen(1999)]{nei99} Neilsen, E. H., Jr. 1999, Ph.D. Thesis, Johns Hopkins Univ.

\bibitem[Olszewski et al.(1993)]{ops93} Olszewski, E.~W.,
Pryor, C., \& Shommer, R.~B.\ 1993, The Globular Cluster-Galaxy Connection,
48, 99

\bibitem[Peng et al.(2004a)]{pen04a} Peng, E. W., Ford, H. C., \&  Freeman, K. C. 2004a, \apjs, 150, 367 
\bibitem[Peng et al.(2004b)]{pen04b} Peng, E. W., Ford, H. C., \&  Freeman, K. C. 2004b, \apj, 602, 705  
\bibitem[Peng et al.(2006)]{pen06} Peng, E.~W., et al. 2006, \apj, 639, 95 

\bibitem[Peterson(1985)]{pet85} Peterson, R.~C.\ 1985, \apj,
297, 309

\bibitem[Peterson et al.(1986)]{poa86} Peterson, R.~C.,
Olszewski, E.~W., \& Aaronson, M.\ 1986, \apj, 307, 139

\bibitem[Peterson et al.(1990)]{pet90} Peterson, R.~C.,
Kurucz, R.~L., \& Carney, B.~W.\ 1990, \apj, 350, 173

\bibitem[Pierce et al.(2006)]{pie06} Pierce, M., et al. 2006, \mnras, 368, 325
\bibitem[Pinkney et al.(2003)]{pin03} Pinkney, J., et al. 2003, \apj, 596, 903
\bibitem[Randall et al.(2004)]{ran04} Randall, S.W., Sarazin, C.L.,
\& Irwin, J.A. 2004, \apj, 600, 729
\bibitem[Randall et al.(2006)]{ran06} Randall, S.W., Sarazin, C.L.,
\& Irwin, J.A. 2006, \apj, 636, 200
\bibitem[Richtler et al.(2004)]{ric04} Richtler, T., et al. 2004, \aj, 127, 2094
\bibitem[Sandage \& Bedke(1994)]{san94} Sandage, A., \& Bedke, J. 1994,
The Carnegie Atlas of Galaxies (Washington: Carnegie Inst.)
\bibitem[Sarazin et al.(2003)]{sar03} Sarazin, C. L., Kundu, A., Irwin, J. A.,
  Sivakoff, G. R., Blanton, E. L., Randall, S. W., 2003, \apj, 595, 743
\bibitem[Schlegel et al.(1998)]{sch98} Schlegel, D.J., Finkbeiner,
D.P., \& Davis, M. 1998, \apj, 500, 525

\bibitem[Schuberth et al.(2006)]{sch06} Schuberth, Y., Richtler, T., Dirsch, B.,
Hilker, M., Larsen, S. S., Kissler-Patig, M., \& Mebold, U. 2006,
\aap, 459, 391 
\bibitem[Shetrone(1996)]{she96} Shetrone, M.~D.\ 1996, \aj,
112, 1517
\bibitem[Sneden et al.(1991)]{sne91} Sneden, C., Kraft,
R.~P., Prosser, C.~F., \& Langer, G.~E.\ 1991, \aj, 102, 2001

\bibitem[Soderberg et al.(1999)]{sod99} Soderberg, A.~M.,
Pilachowski, C.~A., Barden, S.~C., Willmarth, D., \& Sneden, C.\ 1999,
\pasp, 111, 1233

\bibitem[Strader et al.(2006)]{str06} Strader, J., Brodie, J. P., Spitler, L., \&
Beasley, M. A. 2006, \aj, 132, 233 

\bibitem[Suntzeff et al.(1988)]{sun88} Suntzeff, N.~B.,
Kraft, R.~P., \& Kinman, T.~D.\ 1988, \aj, 95, 91

\bibitem[Tonry \& Davis(1979)]{ton79} Tonry, J., \& Davis, M. 1979, \aj, 82, 954

\bibitem[Webbink(1981)]{web81} Webbink, R.~F.\ 1981, \apjs,
45, 259

\bibitem[Woodley et al.(2007)]{woo07} Woodley, K.~A., Harris, 
W.~E., Beasley, M.~A., Peng, E.~W., Bridges, T.~J., Forbes, D.~A., \& 
Harris, G.~L.~H.\ 2007, \aj, 134, 494 

\bibitem[Zepf et al.(2000)]{zep00} Zepf, S.E., Beasley, M.A., Bridges, T.J.,
Hanes, D.A., Sharples, R.M., Ashman, K.M., \& Geisler, D. 2000, \aj, 120, 2928

\bibitem[Zinn \& West(1984)]{zw84} Zinn, R., \& West, M.~J.\ 1984, \apjs, 55, 45

\bibitem[Zinn(1985)]{zinn85} Zinn, R.\ 1985, \apj, 293, 424


\end{thebibliography}
\end{document}